%% file: spf.tex
\RequirePackage{fix-cm}
\documentclass[twocolumn]{svjour3}
\smartqed
\usepackage{latexsym}
\usepackage{amsmath}
\usepackage{amssymb}
\usepackage{braket}
\usepackage{color}
\usepackage{widetext}
\usepackage{graphicx}
\usepackage{mathptmx}
\usepackage{mcite}

\newcommand{\gV}[1]{\boldsymbol #1}
\newcommand{\V}[1]{{\bf #1}}
\journalname{}
\begin{document}

\title{The Semiclassical Propagator in Fermionic Fock Space
}

\author{Thomas Engl \and Peter Pl\"o\ss l \and Juan Diego Urbina \and Klaus Richter
}

\authorrunning{T.~Engl, P.~Pl\"o\ss l, J.~D.~Urbina and K.~Richter}

\institute{T.~Engl \at
              Institut f\"ur Theoretische Physik, Universit\"at Regensburg, D-93040 Regensburg, Germany \\
              \email{thomas.engl@physik.uni-regensburg.de} 
}

\date{}

\maketitle

\begin{abstract}
We present a rigorous derivation of a semiclassical propagator for anticommuting (fermionic) degrees of freedom, starting from an exact representation in terms of Grassmann variables. As a key feature of our approach the anticommuting variables are integrated out exactly, and an exact path integral representation of the fermionic propagator in terms of commuting variables is constructed. Since our approach is \emph{not} based on auxiliary (Hubbard-Stratonovich) fields, it surpasses the calculation of fermionic determinants yielding a standard form $\int {\cal D}[\psi,\psi^{*}] {\rm e}^{i R[\psi,\psi^{*}]}$ with real actions for the propagator. These two features allow us to provide a rigorous definition of the classical limit of interacting fermionic fields and therefore to achieve the long-standing goal of a theoretically sound construction of a semiclassical van Vleck-Gutzwiller propagator in fermionic Fock space. As an application, we use our propagator to investigate how the different universality classes (orthogonal, unitary and symplectic) affect generic many-body interference effects in the transition probabilities between Fock states of interacting fermionic systems.
\keywords{Path integral \and Semiclassical \and Fermions \and classical limit}

\end{abstract}
\input{introduction.tex}
\input{coherent_states.tex}
\input{path_integral.tex}
\input{semiclassical.tex}
\input{transition_probability.tex}
\input{conclusion.tex}
\begin{acknowledgements}
We thank T.~Guhr, P.~Schlagheck, S.~Essert and S.~Smirnov for useful discussions. This work was financially supported by the Deutsche Forschungsgemeinschaft wihtin FOR 760 and SPP 1666.
\end{acknowledgements}
\appendix
\input{appendix.tex}
\bibliographystyle{spphys}
\bibliography{spf}

\end{document}

%% file: introduction.tex
\section{Introduction}
\label{sec:intro}
Semiclassical techniques attempt to describe quantum phenomena using only classical information as input (besides $\hbar$), but keeping all the kinematical and interpretational aspects of quantum mechanics untouched. Semiclassical methods should therefore be distinguished from quasi-classical approaches, which are based on the quantum-classical correspondence and do not only use classical information, but also try to export classical concepts to approximate quantum mechanics. The epitome of the quasi-classical approach is the use of the Ehrenfest theorem to approximate the quantum mechanical evolution of wave packets, with systematic corrections given by the Wigner-Moyal expansion \cite{Brack+Bhaduri}.

Semiclassical methods, as understood in this contribution, attempt to link classical and quantum mechanics in a more abstract, less direct way. While for the quasi-classical program, quantum mechanics is used to construct quantities with a direct classical counterpart (like the trajectory defined by the mean position and momentum of a wavepacket), the semiclassical program employs information extracted from classical trajectories (like their actions and stabilities) to construct quantum mechanical objects. This difference becomes very explicit when we use semiclassical methods to construct quantum objects without classical analogue, such as probability amplitudes.

A major goal of the semiclassical program is the construction of the semiclassical propagator $K^{{\rm sc}}$, the asymptotic form (when $\hbar \to 0$) of the quantum mechanical propagator
\begin{equation}
K({\bf q},{\bf q}^\prime,t)=\langle {\bf q}|{\rm e}^{-\frac{i}{\hbar}\hat{H}t}|{\bf q}^\prime \rangle,
\end{equation}
defined as the matrix element of the time-evolution operator \cite{Sakurai}.

As reviewed in \cite{Gutzwiller}, the challenge to construct a semiclassical propagator has a long history. Although already in 1926 it was clear for Pauli, Dirac and van Vleck that the quantum mechanical propagator can be approximated by an object of the form $K^{{\rm sc}}\sim {\rm e}^{\frac{i}{\hbar}R}$ with the classical action $R$ appearing as a phase, it took more than forty years before Gutzwiller completed the rigorous construction of the semiclassical propagator from Feynman's path integral \cite{Schulmann}. In its final form it reads \cite{Gut1970}
\begin{equation}
K^{\rm sc}({\bf q},{\bf q}^\prime,t)=\sum_{\gamma}A_{\gamma}({\bf q},{\bf q}^\prime,t){\rm e}^{\frac{i}{\hbar}R_{\gamma}({\bf q},{\bf q}^\prime,t)+i\mu_{\gamma}\frac{\pi}{2}}
\label{eq:vvG-propagator_first_quantization}
\end{equation}
where the sum extends over the set of solutions $\gamma$ of the classical problem to join the classical configurations ${\bf q}^\prime$ and ${\bf q}$ in time $t$. As envisioned by Dirac, $R$ is the classical action of the trajectory, while $A$ is related to its variations with respect to the initial and final configurations, and $\mu$ is the number of focal points of the trajectory $\gamma$. 

The derivation of the  van Vleck-Gutzwiller propagator marks the starting point of modern semiclassical methods \cite{Brack+Bhaduri,Gutzwiller,Haake}. They have been able not only to capture but also to successfully describe interference phenomena, \textit{i.e.}~wave effects impossible to describe using quasi-classical techniques.

By Fourier-transforming $K^{{\rm sc}}$ we get the semiclassical (Gutzwiller) Green's function, the starting point to describe stationary properties of quantum systems in the semiclassical limit, and in particular to understand the emergence of universal fluctuations in the spectra and eigenfunctions of classically chaotic quantum systems \cite{Gutzwiller,Haake}. Also, the early semiclassical notion of the theory of molecular collisions \cite{molecular_collisions} and related approaches in mesoscopic condensed matter to describe quantum transport \cite{weaklocalizationBaranger,quantumchaostransportKlaus} (for reviews see \cite{Klaus,Rodolfo,bookdaniel}) connect the van-Vleck propagator, or the semiclassical Green function, with the single-particle S-matrix in terms of transition amplitudes for transmission and reflection.

The success of the semiclassical methods has been restricted, however, predominantly to quantum systems that admit a first-quantization description. In fact, the generalization of the van Vleck-Gutzwiller propagator to describe systems of interacting particles does not pose any conceptual challenge, as the classical limit of the theory is very well understood. The semiclassical propagator is now an established tool to describe quantum dynamics of molecular systems \cite{miller_s-matrix,cellular_dynamics,nonadiabatic,ivr} and mesoscopic electronic systems \cite{Ullmo}.

Technical, but not conceptual, problems arise when indistinguishability comes into play. Here, the semiclassical calculation of ground and (doubly) excited states in helium by Greg Ezra et.~al.~\cite{helium_semiclassical} marks a successful step in coping with strongly interacting two-electron dynamics. The number of classical paths we need to construct to calculate the transition amplitude between different (anti-) symmetrized configurations of a quantum system however grows extremely fast with the number of particles \cite{Quirin}. The same vast increase of the number of classical trajectories that have to be taken into account, affects the coupled coherent state approach \cite{ccs}, which has been developed for the treatment of fermionic many-body systems in phase space. In this approach, the wave function is expanded in a (large) set of Slater determinants of single-particle coherent states with randomly selected intitial conditions. The coherent states are then evolved along the corresponding classical trajectory.

Moreover, for fermionic systems with spin orbit interactions, hybrid semiclassical approaches exist, which describe the orbital motion of non-interacting particles in phase space, while the spin is treated in a second quantized approach using spin coherent states \cite{Paul,semiclassical_soi_strong_coupling1,*semiclassical_soi_strong_coupling2,semiclassical_soi_transport,Bolte2001,semiclassical_soi_trace-formula1,semiclassical_soi_trace-formula2,semiclassical_weak_antilocalization}

Importantly, the emergence of mean-field behavior, an expected simplification of the description when the number of particles is large, cannot be rigorously included in a natural way if one sticks to the first-quantized picture where the total number of particles $N$ is not defined by the quantum many body state but is an external parameter determining the dimensionality $D=Nd$, where $d$ is the spatial dimension, of the system and thereby fixing the structure of the very space where the system lives.

These remarks indicate already a possible solution of the problem. If a second-quantized picture in Fock space is adopted instead, both quantum indistinguishability and flexibility in the number of particles are automatically included at the kinematic level: the Fock space of quantum states is by definition spanned by states which are correctly (anti-) symmetrized, and the number of particles is simply another observable represented by a hermitian operator \cite{negele+orland}. When invoking a Fock space description, this change of perspective implies for the semiclassical program that particles appear as an emergent concept, derived from the more fundamental degree of freedom: the quantum field \cite{Weinberg}.

The development of a semiclassical program for bosonic fields has received powerful impact from the experimental realization of their discrete version in the context of cold-atom physics \cite{Lewenstein}. In fact, the theoretical model that describes microscopically a system of interacting bosons on a lattice, the so-called Bose-Hubbard model \cite{coldatoms-review}, is a special realization of an interacting bosonic field. Here, again, the complementarity between quasi-classical and semiclassical approaches has been apparent. Quasi-classical methods as the ones used in \cite{Aguiar+Barangar} work well as long as quantum interference does not come into play and eventually dominates the dynamics. However, a rigorous derivation of the van Vleck-Gutzwiller propagator in bosonic Fock space was achieved only recently \cite{cbs_fock}.

It is fair to say that the situation in the fermionic case is more desperate. Already a quasi-classical approach faces a fundamental problem: how to define a sensible classical limit if the fermionic fields must obey the Pauli principle and therefore admit only non-commutative descriptions? The attempts and achievements to associate commuting variables to fermionic operators, that spans from the 1970's well into the 2010's, are still lacking a rigorous microscopic derivation, indicating the complexity of the problem \cite{Klauder,miller_classical_spin-analogy,miller_classical,Grossmann}. The importance of the Chemical Physics community in this program has been obvious: electronic degrees of freedom are fundamental in the realm of molecular reactions. Moreover, chemical reactions require, in principle, simulations with anticommuting variables.

In order to avoid these anticommuting variables, in a series of important papers, Miller and collaborators proposed to use a heuristic generalization of the Heisenberg prescription \cite{miller_classical_spin-analogy,miller_classical,miller_semiclassical} to construct the classical limit of fermionic degrees of freedom (for recent applications see \cite{Anderson_impurity,molecular_transport}). It is a remarkable and valuable feature of this approach that it associates correct signs to expressions involving anticommuting fermionic operators $\hat{c},\hat{c}^{\dagger}$ and respects the Pauli principle. In the simplest example, these key features can be seen in the mapping $F \to F^{{\rm cl}}$ between operators $F(\hat{c},\hat{c}^{\dagger})$ and classical phase space functions $F^{{\rm cl}}(\sqrt{n}{\rm e}^{i\theta},\sqrt{n}{\rm e}^{-i\theta})$, which gives for $i \ne j$
\begin{eqnarray}
\label{eq:Miller}
\hat{c}_{i}^{\dagger}\hat{c}_{j} &\to& \sqrt{n_{i}n_{j}(1-n_{i})(1-n_{j})}{\rm e}^{-i(\theta_{i}-\theta_{j})}, \\
\hat{c}_{j}\hat{c}^{\dagger}_{i} &\to& -\sqrt{n_{i}n_{j}(1-n_{i})(1-n_{j})}{\rm e}^{-i(\theta_{i}-\theta_{j})}. \nonumber
\end{eqnarray}
The (in general continuous) classical phase space variables $0 \le n \le 1$ are naturally interpreted as classical fermionic occupation numbers with the angles $\theta$ as their corresponding canonically conjugated variables.

However, as it is obvious from eq.~(\ref{eq:Miller}), the thus  classical Hamiltonian obtained in this way has the physical Fock states, defined by
\begin{equation}
n_{i}=0 {\rm \ or \ } 1 {\rm \ for \ all \ } i,
\end{equation}
as fixed points of the dynamics and the corresponding semiclassical propagator is then trivially incorrect in the relevant case where it connects physical Fock states. Moreover, as discussed at length in Sec.~\ref{sec:path_integral}, approaching the classical limit from the quantum side by means of a formal path integral in terms of the fermionic states (introduced by Klauder \cite{Klauder}),
\begin{equation}
|b\rangle=b|0\rangle +\sqrt{1-|b|^{2}}\hat{c}^{\dagger}|0\rangle {\rm \ \ , \ \ with \ complex \ } b,
\end{equation}
shows that eq.~(\ref{eq:Miller}) can be rigorously obtained from an exact path integral representation in terms of the commuting fields $b$. This indicates that in a representation where eq.~(\ref{eq:Miller}) holds, the quantum mechanical propagation between Fock states is {\it not} supported by classical trajectories and the semiclassical limit is problematic.   

This complication may be due to the fact that in Klauder's representation the path integral is {\it restricted}, namely, the integration over the variables $b$ are defined inside the unit disk instead of over the whole complex plane. A heuristic incorporation of Langer corrections proposed in \cite{miller_semiclassical},
\begin{equation}
\sqrt{n(1-n)} \to \sqrt{\left(n+\frac{1}{2}\right)\left(\frac{3}{2}-n\right)},
\end{equation}
lifts the problem and actually leads to a classical limit that gives, for example, agreement with first-order quantum perturbation theory by using classical perturbation theory.

As this volume commemorates Greg Ezra's contributions to the description of atomic and molecular dynamics, we would like to mention that Ezra's pioneering work on the Langer correction to the semiclassical propagator \cite{Ezra} could possibly provide the key to make rigorous the promising proposal presented in \cite{miller_classical}. It is then tempting to check whether Ezra's insight into Langer corrections within the path integral formalism in first-quantized systems with would help to make Miller's approach justified from first-principles \cite{us}.

Here we follow a different route and present what we believe to be the first microscopic derivation of the exact propagator between $N$-particle fermionic Fock states in terms of path integrals over commuting, unrestricted classical fields. Our path integral not only incorporates and generalizes Miller's mapping $F \to F^{{\rm cl}}$ "teaching" the classsical limit of large $N$ about anticommuting operators, but it is supported in the semiclassical limit by classical paths. No extra assumptions or corrections are required.

As we will discuss in Sec.~\ref{sec:path_integral}, the thus derived classical Hamiltonian corresponds to an approximation of the Holstein-Primakoff transformation for a single particle in a two-level system, used in \cite{mmst-mapping3}.

After briefly introducing Grassmann variables in Sec.~\ref{sec:coherent_states}, in Sec.~\ref{sec:path_integral} we present our derivation of the exact path integral for fermionic systems. Armed with this object, in Sec.~\ref{sec:semiclassical_approximation} we follow the typical semiclassical program: we identify both the effective Planck's constant and the classical limit of the theory from the phase of the path's amplitude in the path integral, and evaluate the path integral in stationary phase approximation to obtain a van Vleck-Gutzwiller type propagator for interacting fermionic fields. The presentation will be restricted to spin-$1/2$ systems, although a generalization to higher spins is straight forward. Finally, in Sec.~\ref{sec:transition_prob}, we use the thus derived semiclassical propagator to calculate the transition probability from one fermionic Fock state to another one for systems without time reversal symmetry, for systems diagonal in spin space but time reversal invariant, as well as for time reversal invariant spin-$1/2$ systems non-diagonal in spin space. 

Technical details of the derivation of our main results, namely the exact complex path integral representation of the fermionic propagator in terms of commuting fields, eq.~(\ref{eq:propagator_in_complex_path-integral}), the classical Hamiltonian eq.~(\ref{eq:classical_hamiltonian}) and the van Vleck propagator, eqns.~(\ref{eq:semiclassical_propagator_before_fluctuation-integration},\ref{eq:semiclassical_prefactor}) can be found in the appendices.

%% file: coherent_states.tex
\section{Grassmann coherent states}
\label{sec:coherent_states}
In order to derive the path integral representation for the fermionic propagator in Fock space, we will use Grassmann coherent states in intermediate steps. They are defined as the eigenstates of the fermionic annihilation operators \cite{negele+orland},
\begin{equation}
\hat{c}_j\ket{\gV{\zeta}}=\zeta_j\ket{\gV{\zeta}}.
\label{eq:definition_coherent_states}
\end{equation}
Here, $\hat{c}_j$ and $\hat{c}_j^\dagger$ annihilates and creates, respectively, a particle in the $j$-th single particle state, two states which coincide in the orbital degrees of freedom, but differ in the spin degree of freedom are accounted for as different single particle states, and are therefore labeled by different indexes $j$. 

However, due to the antisymmetry and the Pauli exclusion principle, the eigenvalues of the coherent states have to be (complex) anticommuting numbers, called Grassmann numbers \cite{negele+orland,grassmann_numbers}, \textit{i.e.}~for two of these numbers $\zeta$ and $\chi$
\begin{equation}
\zeta\chi=-\chi\zeta.
\label{eq:anticommutativity_grassmann}
\end{equation}
They also anticommute with the creation and annihilation operators,
\begin{align}
\zeta\hat{c}_j^{}&=-\hat{c}_j^{}\zeta, & \zeta\hat{c}_j^{\dagger}&=-\hat{c}_j^{\dagger}\zeta,
\label{eq:anticommutation_grassman_creation_annihilation}
\end{align}
while they commute with regular complex numbers. The anticommuting property also implies $\zeta^2=0$.

Integration over a complex Grassmann number is defined by
\begin{align}
&\int{\rm d}\zeta^\ast{\rm d}\zeta1=\int{\rm d}\zeta^\ast{\rm d}\zeta\zeta=\int{\rm d}\zeta^\ast{\rm d}\zeta\zeta^\ast=0,
\label{eq:vanishing_grassmann-integrals} \\
&\int{\rm d}\zeta^\ast{\rm d}\zeta\zeta\zeta^\ast=1.
\label{eq:nonvanishing_grassmann-integrals}
\end{align}
With the properties of the Grassmann numbers, it is possible to show that the fermionic coherent states are given by \cite{negele+orland}
\begin{equation}
\ket{\gV{\zeta}}=\exp\left(-\frac{1}{2}\gV{\zeta}^{\ast}\cdot\gV{\zeta}\right)\prod\limits_{j}\left(1-\zeta_j^{}\hat{c}_j^\dagger\right)\ket{0},
\label{eq:construction_coherent_states}
\end{equation}
where $\ket{0}$ denotes the fermionic vacuum state. Moreover, they satisfy
\begin{align}
&\Braket{\gV{\zeta}|\gV{\chi}}=\exp\left[\sum\limits_{j}\left(-\frac{1}{2}\zeta_j^\ast\zeta_j^{}-\frac{1}{2}\chi_j^\ast\chi_j^{}+\zeta_j^\ast\chi_j^{}\right)\right],
\label{eq:overlap_coherent-coherent} \\
&\Braket{\V{n}|\gV{\zeta}}=\exp\left(-\frac{1}{2}\gV{\zeta}^{\ast}\cdot\gV{\zeta}\right){\prod\limits_{j}}^\prime\zeta_j^{n_j}, \\
&\int{\rm d}\gV{\zeta}^\ast\int{\rm d}\gV{\zeta}\ket{\gV{\zeta}}\bra{\gV{\zeta}}=1,
\label{eq:overlap_coherent-fock}
\end{align}
with $\ket{\V{n}}$ being an arbitrary Fock state, such that $n_j\in\{0,1\}$ is the occupation of the $j$-th single particle state. The prime at the product indicates that the order of the individual factors is reversed, \textit{i.e.}~the factor corresponding to the largest possible value is the most left one, while the $j=1$ term is the most right one.

%% file: path_integral.tex
\section{The path integral in complex variables}
\label{sec:path_integral}
\subsection{Derivation}
The aim of this part is to derive a path integral representation of the propagator in Fock space,
\begin{equation}
K\left(\V{n}^{(f)},\V{n}^{(i)};t_f\right)=\Braket{\V{n}^{(f)}|\exp\left(-\frac{\rm i}{\hbar}\hat{H}t_f\right)|\V{n}^{(i)}},
\label{eq:definition_propagator_fock}
\end{equation}
to which the stationary phase approximation can be applied. Note that for simplicity of presentation, the Hamiltonian has been chosen time independent, although the following calculations are also valid for the time dependent case.

The path integral representation is usually achieved by applying the Trotter Formula \cite{trotter}, which replaces the exponential in eq.~(\ref{eq:definition_propagator_fock}) by the product of infinitely many propagators with an infinitesimally small time step and by inserting the unit operator between two adjacent factors. Since the resolution of unity for Fock states is given by a sum, rather than an integral, they are not suitable for the construction of a path integral. This makes the coherent states the natural choice for the representation of the unit operator. However, when applying the semiclassical approximation to the coherent state path integral, one ends up with grassmannian equations of motion. On the other hand, it is desirable to have complex equations of motion leading to a real action. In order to achieve this, one has to find a way to replace the integrals over Grassmann variables by integrals over complex ones.

Here, we will give a rough description of the procedure, which allows for such a transformation from Grassmann to complex integrals. However, it turns out that some of the steps contain a certain freedom of choice. The final path integral will then depend on the individual choices made during the derivation. The derivation for the specific choice presented later in this publication, is then carried out in appendix \ref{app:pathintegral}.

After applying Trotter's formula \cite{trotter} the first step is to insert \emph{two} unit operators in terms of fermionic coherent states between two adjacent exponentials,
\begin{align}
K&\left(\V{n}^{(f)},\V{n}^{(i)};t_f\right)= \nonumber \\
&\lim\limits_{M\to\infty}\left[\prod\limits_{m=0}^{M}\left(\int{\rm d}{\gV{\zeta}^{(m)}}^{\ast}\int{\rm d}{\gV{\zeta}^{(m)}}^{}\int{\rm d}{\gV{\chi}^{(m)}}^{\ast}\int{\rm d}{\gV{\chi}^{(m)}}^{}\right)\right] \nonumber \\
&\quad\left[\prod\limits_{m=0}^{M-1}\Braket{\gV{\zeta}^{(m+1)}|\exp\left(-\frac{{\rm i}\tau}{\hbar}\hat{H}\right)|\gV{\chi}^{(m)}}\Braket{\gV{\chi}^{(m)}|\gV{\zeta}^{(m)}}\right] \nonumber \\
&\quad\Braket{\V{n}^{(f)}|\gV{\chi}^{(M)}}\Braket{\gV{\chi}^{(M)}|\gV{\zeta}^{(M)}}\Braket{\gV{\zeta}^{(0)}|\V{n}^{(i)}},
\label{eq:propagator_in_grassmann_path-integral}
\end{align}
where $\tau=t_f/M$.

Next, in order to replace the Grassmann integrals by complex ones, one has to insert complex integrals such that the overlap $\braket{\gV{\chi}^{(m)}|\gV{\zeta}^{(m)}}$ can be written as an integral over a product of two factors, with the first one depending only on $\gV{\chi}^{(m)}$ and the second one on $\gV{\zeta}^{(m)}$. Here, integrals of the form
\begin{equation}
\int\limits_{\mathbb{C}}^{}{\rm d}\phi\int\limits_{\mathbb{C}}^{}{\rm d}\mu\exp\left(-\left|\phi\right|^2-\left|\mu\right|^2+\phi^{\ast}\mu\right)\phi^{k^{}}\left(\mu^{\ast}\right)^{k^{\prime}}=\pi^2k!\delta_{k^{}k^\prime}
\label{eq:inserted_integral}
\end{equation}
will be used, since this choice allows us to construct a path integral, which for intermediate times has the same form as the one for bosons in coherent state representation \cite{negele+orland} (see appendix \ref{app:pathintegral}).

After this insertion, we can decouple $\gV{\zeta}^{(m+1)}$ and $\gV{\chi}^{(m)}$ from $\gV{\zeta}^{(m)}$ and $\gV{\chi}^{(m-1)}$ in eq.~(\ref{eq:propagator_in_grassmann_path-integral}), such that the integrand for the propagator becomes a product, in which the $m$-th factor only depends on $\gV{\zeta}^{(m)}$ and $\gV{\chi}^{(m-1)}$. Therefore the insertion of these integrals allows us to integrate out the Grassmann variables exactly after expanding the exponential up to linear order in $\tau$.

At this point, it is important to note that not only the choice of the inserted integrals is not unique, but that, when choosing \textit{e.g.}~integrals of the form (\ref{eq:inserted_integral}), there is a certain freedom in choosing the combinations of $k$ and $k^\prime$. With the choices cf.~appendix \ref{app:pathintegral}, one arrives at

\begin{widetext}
\begin{align}
K\left(\V{n}^{(f)},\V{n}^{(i)};t_f\right)=\left[\prod\limits_{j:n_j^{(i)}=1}^{}\int\limits_{0}^{2\pi}\frac{{\rm d}\theta_j^{(0)}}{2\pi}\exp\left(-{\rm i}\theta_j^{(0)}\right)\right]\left(\prod\limits_{j:n_j^{(f)}=1}^{}\int\limits_{\mathbb{C}}^{}\frac{{\rm d}\phi_j^{(M)}}{\pi}\phi_j^{(M)}\right)\left(\prod\limits_{m=1}^{M-1}\prod\limits_{j}^{}\int\limits_{\mathbb{C}}^{}\frac{{\rm d}\phi_j^{(m)}}{\pi}\right)\times& \nonumber \\
\times\exp\left\{\sum\limits_{m=1}^{M}\left[-\left|\gV{\phi}^{(m)}\right|^2+{\gV{\phi}^{(m)}}^\ast\cdot\gV{\phi}^{(m-1)}-\frac{{\rm i}\tau}{\hbar}H_{cl}\left({\gV{\phi}^{(m)}}^{\ast},\gV{\phi}^{(m-1)}\right)\right]\right\}&,
\label{eq:propagator_in_complex_path-integral}
\end{align}
\end{widetext}

where at final time the integrals over those $\phi_j^{(M)}$ corresponding to empty single particle states, \textit{i.e.}~for those $j$ where $n_j^{(f)}=0$, are already evaluated exactly and therefore have to be set to zero in eq.~(\ref{eq:propagator_in_complex_path-integral}). In fact, the integrals over those components do not even have to be inserted right from the beginning, since
\begin{equation}
\int{\rm d}{\chi_j^{(M)}}^\ast\int{\rm d}\chi_j^{(M)}\exp\left(-{\chi_j^{(M)}}^\ast\chi_j^{(M)}\right)\left(1+{\chi_j^{(M)}}^\ast\zeta_j^{(M)}\right)=1.
\end{equation}
The exact integration over the finally unoccupied states is necessary, since the stationarity conditions will not give solutions for the phases of these components and therefore, these integrals can not be performed in a stationary phase approximation. For the same reason the integrations over those $\phi_j^{(0)}$ with $n_j^{(i)}=0$ are already performed exactly. This means that effects due to vacuum fluctuations \cite{Nolting}, \textit{i.e.}~the spontaneous creation and annihilation of particles out of the vacuum, are treated exactly. Furthermore, for $m=0$, the integrations over the amplitudes $J_j^{(0)}=|\phi_j^{(0)}|^2$ for the initially occupied single particle states $j$ are performed exactly (see appendix \ref{app:pathintegral} for details of this exact integration). As a matter of fact, these integrals could also be included in the stationary phase approximation, which would eventually result in a multiplication of our result for the semiclassical propagator with a factor $\alpha={\rm e}^N/(\sqrt{2\pi})^N$, where $N$ is the total number of particles, which is the $N$-th power of Stirling's approximation of $n!$ for $n=1$.

Now one might raise the question, why the initial amplitudes related to occupied states are integrated out, but not the final ones. Actually, the amplitudes of $\phi_j^{(M)}$ for occupied sites could also be integrated out, which would result in dividing the result for the semiclassical approximation by the same factor $\alpha$. However, we choose not to perform them, in order to be in accordance with the usual first quantized semiclassical approach, where the path integral, to which the stationary phase approximation is applied, consists of one integration (over the canonical variables chosen as basis) less than those over their canonical conjugate variables. For instance, the path integral for the propagator in configuration space consists (before taking the limit $M\to\infty$) of $M$ momentum integrals and $M-1$ position integrals. Moreover, our choice is supported by the fact that it leads to the exact result if the quantum Hamiltonian is diagonal and non-interacting.

When comparing the path integral with the corresponding one in first quantization, eq.~(\ref{eq:vvG-propagator_first_quantization}), the phases $\theta_j^{(0)}$ would correspond to the initial momenta of the path. The role of $\gV{\phi}^{(M)}$, however, is much more sophisticated. Its phases again correspond to the final momenta, while its amplitude should somehow correspond to the final position. Yet, the value of the latter is not fixed to $n_j^{(f)}=1$, which would be the expected boundary condition for the paths. This boundary condition is hidden in the in the integration over $\phi_j^{(M)}$ is determined by the extra factor $\phi_j^{(M)}$ of the integrand. In a stationary phase analysis of the integrand, which will be performed below, one finally recognizes that indeed both, the stationarity condition of phase and amplitude of $\phi_j^{(M)}$, are required in order to get the correct boundary condition. Thus, the boundary condition at final time is indeed hidden in the full integral over $\phi_j^{(M)}$.

Finally, it should be noted that the classical Hamiltonian $H_{cl}$ is not unique, but again depends on the way chosen to construct the path integral in complex variables. There remains a certain freedom to weigh individual terms in the classical Hamiltonian differently, which might help in studying effects related to particular parts of the Hamiltonian. For instance, in the Hamiltonian given in eq.~(\ref{eq:classical_hamiltonian_2}) in appendix \ref{subapp:particle-picture}, the interaction, single-particle energies and the antisymmetry under particle exchange are weighted exponentially, while the Pauli principle is given by an exponential suppression of hopping processes leading to occupations of one single-particle state by more than one particle. However, due to the exponential factor in the diagonal term of the single-particle part of the Hamiltonian, processes quantum mechanically forbidden by the Pauli principle are further suppressed energetically. This energetically suppression essentially corresponds to the heuristic inclusion of a Pauli potential \cite{Grossmann,paulipotential1,paulipotential2,paulipotential3,paulipotential4}, \textit{i.e.}~a potential, which hinders two electrons to occupy the same single-particle state.

For the quantum Hamiltonian considered here,
\begin{equation}
\hat{H}=\sum\limits_{\alpha,\beta}h_{\alpha\beta}\hat{c}_{\alpha}^\dagger\hat{c}_{\beta}^{}+\sum\limits_{\substack{\alpha,\beta \\ \alpha\neq\beta}}^{}U_{\alpha\beta}\hat{c}_{\alpha}^\dagger\hat{c}_{\beta}^\dagger\hat{c}_{\beta}^{}\hat{c}_{\alpha}^{}.
\label{eq:qm_hamiltonian}
\end{equation}
one possible classical Hamiltonian is given by
\begin{align}
H_{cl}&\left(\gV{\mu},\gV{\phi}\right)= \nonumber \\
\sum\limits_{\alpha}&h_{\alpha\alpha}\mu_{\alpha}\phi_{\alpha}+\sum\limits_{\substack{\alpha,\beta \\ \alpha\neq\beta}}U_{\alpha\beta}\mu_{\alpha}\mu_{\beta}\phi_{\alpha}\phi_{\beta}
\label{eq:classical_hamiltonian} \\
+&\sum\limits_{\substack{\alpha,\beta \\ \alpha\neq\beta}}h_{\alpha\beta}\mu_{\alpha}\phi_{\beta}\exp\left(-\mu_{\alpha}\phi_{\alpha}-\mu_{\beta}\phi_{\beta}\right){\prod\limits_{j}}^{\alpha,\beta}\left(1-2\mu_j\phi_j\right), \nonumber
\end{align}
where the product in the last line runs only over those values of $j$, which are lying between $\alpha$ and $\beta$, excluding $\alpha$ and $\beta$ themselves. The case $\gV{\mu}=\gV{\phi}^{*}$, \textit{i.e.}
\begin{align}
H_{cl}&\left(\gV{\phi}^{*},\gV{\phi}\right)= \nonumber \\
\sum\limits_{\alpha}&h_{\alpha\alpha}|\phi_{\alpha}|^{2}+\sum\limits_{\substack{\alpha,\beta \\ \alpha\neq\beta}}U_{\alpha\beta}|\phi_{\alpha}|^{2}|\phi_{\beta}|^{2}
\label{eq:classical_hamiltonian_diagonal} \\
+&\sum\limits_{\substack{\alpha,\beta \\ \alpha\neq\beta}}h_{\alpha\beta}\phi_{\alpha}^{*}\phi_{\beta}\exp\left(-|\phi_{\alpha}|^{2}-|\phi_{\beta}|^{2}\right){\prod\limits_{j}}^{\alpha,\beta}\left(1-2|\phi_{j}|^{2}\right), \nonumber
\end{align}

will be of particular importance for the continuum limit. It is instructive to compare it with the classical electron analog model (CEAM) obtained from Miller's mapping which gives in this case
\begin{align}
H_{cl}^{{\rm CEAM}}&\left(\gV{\phi}^{*},\gV{\phi}\right)= \nonumber \\
\sum\limits_{\alpha}&h_{\alpha\alpha}|\phi_{\alpha}|^{2}+\sum\limits_{\substack{\alpha,\beta \\ \alpha\neq\beta}}U_{\alpha\beta}|\phi_{\alpha}|^{2}|\phi_{\beta}|^{2}
\label{eq:classical_hamiltonian_Miller} \\
+&\sum\limits_{\substack{\alpha,\beta \\ \alpha\neq\beta}}h_{\alpha\beta}\phi_{\alpha}^{*}\phi_{\beta}\sqrt{(1-|\phi_{\alpha}|^{2})(1-|\phi_{\beta}|^{2})}{\prod\limits_{j}}^{\alpha,\beta}\left(1-2|\phi_{j}|^{2}\right), \nonumber
\end{align}
in terms of the, now {\it restricted}, variables $\phi_{\alpha}$ with $|\phi_{\alpha}|^{2} \le 1$.

In eq.~(\ref{eq:classical_hamiltonian}), the factors $1-2\mu_j\phi_j$ are a consequence of the anticommutativity of the creation and annihilation operators (and the Grassmannians) and thus account for the antisymmetry of the fermions under particle exchange. Consider for example the following two processes for the scattering of two particles in the states $1$ and $2$ into the states $2$ and $3$: in the first process, the particle in state $1$ is scattered into state $3$, with the second particle staying in state $2$, while in the second one the particle in state $2$ is scattered into state $3$ and the particle in state $1$ is scattered into state $2$. These two processes are the same up to an exchange of the two particles. Therefore, these two processes have to yield the same contribution, but with a different sign. On the other hand, if state $2$ is empty, while a particle is scattered from state $1$ to state $3$, there is no corresponding process resulting from an odd number of exchanges of particles, and thus, the contribution has always to be the same. In general, a process where a particle is scattered from state $\alpha$ to state $\beta$ with $\left|\alpha-\beta\right|>1$, has to be multiplied by a factor of $-1$ for each occupied state $j$ between $\alpha$ and $\beta$. However, classically the occupations are not restricted to $0$ and $1$, but can be any number, such that one ends up with a factor interpolating between the two extreme values $+1$ for the case without a particle in state $j$ and $-1$ for the case where state $j$ is occupied. Furthermore, the exponential in the non-diagonal part of the single particle term accounts for the Pauli principle by the exponential suppression of processes, which lead to an enhanced number of particles within one single particle state.

A (certainly not complete) list of further possible classical Hamiltonians corresponding to the quantum Hamiltonian (\ref{eq:qm_hamiltonian}) can be found in appendix \ref{app:classical_hamiltonians}.

It is furthermore instructive to see how our approach treats the extreme case of a {\it single} electron, $N=1$, where the state space is spanned by two discrete states and anticommutation of the fermionic fields does not play a role. In this situation, our results can be directly compared with existing exact mappings between systems with $n=2$ discrete states and a quantum top with total angular momentum $s$ such that $n=(2s+1)/2$. In the  Chemical Physics community these so-called Meyer-Miller-Stock-Thoss (MMST) methods \cite{miller_classical_spin-analogy,molecular_transport,mmst-mapping2,mmst-mapping3} have been successfully used to describe non-adiabatic transitions of the nuclear dynamics between two potential surfaces corresponding to two discrete many-body states of the electrons. The MMST method maps the dynamics of a two-level system into the problem of a spinning particle, which can be in turn mapped into a set of harmonic oscillators by means of the Schwinger representation of angular momentum (see \cite{mmst-mapping3}). In this way, a classical picture for two-level systems is obtained, as a basis for standard (continuous) semiclassical approaches.

Our result for the classical limit of a single electron, included in Eq.~(\ref{eq:classical_hamiltonian_diagonal}), appears naturally within the MMST approach as an approximate version of the Holstein-Primakoff transformation, see \cite{mmst-mapping3} for details and \cite{Paul} for an application to spin transport. As it is also shown there, this classical limit, however, gives unsatisfactory results when used as starting point of a semiclassical calculation of the time evolution of quantum observables. This apparent drawback is fully resolved when taking into account, as shown in detail here, that the semiclassical limit where our result holds is defined by $N \to \infty$. Therefore, the application of our methods to the limiting case $N=1$ is expected to poorly compare with exact quantum mechanical results. However, the main motivation of the present work is to deal semiclassically with anticommuting variables, not with few discrete degrees of freedom as in \cite{mmst-mapping3}.
\subsection{Comparison with CEAM and Klauder's approach}
Miller's heuristic approach can actually be verified by extracting the classical Hamiltonian from another path integral representation. This is by extending the $b$-fermionic states introduced by Klauder in \cite{Klauder},
\begin{equation}
\ket{b}=\sqrt{1-\left|b\right|^2}\ket{0}+b\ket{1},
\label{eq:klauder}
\end{equation}
to the case of multiple single-particle states and define (see also \cite{Levein})
\begin{equation}
|{\bf b}\rangle=\prod_{j}\left(\sqrt{1-|b_{j}|^{2}}\hat{1}+b_{j}\hat{c}^{\dagger}_{j}\right)\ket{0}.
\end{equation}
These states define an overcomplete basis for the fermionic Hilbert space, as they form the identity
\begin{equation}
\left(\prod\limits_{j}^{}\int\limits_{\mathbb{D}}^{}\frac{{\rm d}b_j^{(m)}}{\pi}\right)\ket{\bf b}\bra{\bf b}=\hat{1}
\end{equation}
where $\mathbb{D}$ denotes the unit disc in the complex plane, and therefore can be used to construct a path integral representation of the propagator in terms of paths ${\bf b}(t)$ in the space of commuting variables ${\bf b}$.

The steps of the derivation of the path integral in this basis correspond to those one follows to construct the fermionic path integral using coherent states \cite{negele+orland,Klauder}. After reaching a form where the classical Hamiltonian can be read off from an action functional giving the phase of the quantum propagator, we obtain
\begin{equation}
H^{{\rm Klauder}}_{cl}({\bf b}^{*},{\bf b})=\langle {\bf b}|\hat{H}|{\bf b}\rangle.
\label{eq:classical_hamiltonian_Klauder}
\end{equation}
A short calculation finally shows that the classical Hamiltonian (\ref{eq:classical_hamiltonian_Klauder}) obtained using Klauder's representation  is equal to Miller's, eq.~(\ref{eq:classical_hamiltonian_Miller}), \textit{i.e.}
\begin{equation}
H^{{\rm Klauder}}_{cl}({\bf b}^{*},{\bf b})=H^{{\rm CEAM}}_{cl}({\bf b}^{*},{\bf b}).
\end{equation}
thus providing a rigorous construction of the classical limit of the approach by Miller and coworkers \cite{miller_classical}. 

In principle, having at hand a classical Hamiltonian as the one in eq.~(\ref{eq:classical_hamiltonian_Miller}), a semiclassical analysis of the path integral in $b$-representation along the lines presented bellow can be carried out. The first step is to consider the classical equations of motion
\begin{equation}
i\hbar\frac{d}{dt}{\bf b}(t)=\frac{\partial}{\partial {\bf b}^{*}}H^{{\rm CEAM}}_{cl}({\bf b}^{*},{\bf b}),
\end{equation}
which can be canonically transformed into
\begin{align}
i\hbar\frac{d}{dt}n_{j}(t)&=\left.\frac{\partial}{\partial \theta_{j}}H^{{\rm CEAM}}_{cl}({\bf b}^{*},{\bf b})\right|_{b=\sqrt{n}\exp({i\theta})}\\
i\hbar\frac{d}{dt}\theta_{j}(t)&=-\left.\frac{\partial}{\partial n_{j}}H^{{\rm CEAM}}_{cl}({\bf b}^{*},{\bf b})\right|_{b=\sqrt{n}\exp({i\theta})}.
\end{align}
Without loss of generality we consider the many-body Hamiltonian (\ref{eq:qm_hamiltonian}). Inspection of the associated equations of motion readily shows that the classical occupations $n_{j}=|b_{j}|^{2}$ evolve in time only through the terms that depend on the phases $\theta_{j}$. Here is where the classical limit $H^{{\rm CEAM}}_{cl}({\bf b}^{*},{\bf b})$ is problematic: due to the presence of the "Pauli" factors $\sqrt{n(1-n)}$ in eq.~(\ref{eq:classical_hamiltonian_Miller}) we trivially obtain
\begin{equation}
\left.\frac{d}{dt}n_{j}(t)\right|_{n=0 {\rm \ or \ } 1}=0.
\end{equation}
Therefore the classical phase-space manifolds associated with the physical Fock states, which are defined by precisely the condition $n=0 {\rm \ or \ } 1$, do not evolve in time and there is no way to connect the quantum and classical dynamics, neither at the quasi-classical, nor at the semiclassical level. Remarkably, the classical limit as given for example in eq.~(\ref{eq:classical_hamiltonian}) circumvents this problem by allowing arbitrarily high classical occupation numbers, but penalizing them in a smooth (but exponentially strong) manner. 

It is important to stress that there is no reason why classical occupations must be bounded, exactly as there is no reason why they have to take only integer values. In both cases we are apparently violating what is just a classical picture of the fermionic degrees of freedom. However, fermionic fields are essentially non-classical objects and we are satisfied with being able to {\it define} a consistent classical limit by pure formal manipulations. Adopting this pragmatical point of view of defining the classical limit formally through the exact path integral, the fields $\phi_{\alpha}$ in eq.~(\ref{eq:classical_hamiltonian_diagonal}) do not need to fit our expectations on how the classical limit should look like. All that we ask them for is to correctly describe the propagation between physical Fock states.

%% file: semiclassical.tex
\section{Semiclassical approximation}
\label{sec:semiclassical_approximation}

The reason for the semiclassical approach to any quantum system to be rooted in the path integral formulation is that it accomplishes simultaneously three major goals. First, it allows us to identify the classical limit of the theory. Second, it serves as the starting point of a systematic stationary phase analysis that eventually leads to the semiclassical propagator. Third, {\it it is in the structure of the action functional where $\hbar_{{\rm eff}}$ can be identified}. The effective Planck constant is not only the dimensionless parameter that defines the classical limit $\hbar_{{\rm eff}}\to0$, but also the small parameter that makes the whole semiclassical approach valid. It appears non-perturbatively, if the characteristic path integral representation of the propagator,
\begin{equation}
K\sim \int {\cal D}[ \cdot ]{\rm e~}^{R[ \cdot ]/\hbar},
\end{equation}
is written in terms of a dimensionless action $\tilde{R}$,
\begin{equation}
K\sim \int {\cal D}[ \cdot ]{\rm e~}^{\tilde{R}[ \cdot ]/\hbar_{{\rm eff}}}.
\end{equation}
Inspection of the exponents in eq.~(\ref{eq:propagator_in_complex_path-integral}) shows that Planck's constant $\hbar$ actually plays a minor role in our case. Clearly, $\hbar$ can be absorbed simply by a redefinition of the parameters of the Hamiltonian (note that this is not the case in the usual phase-space path integral). In order to identify $\hbar_{{\rm eff}}$, we rescale all the fields in such a way that the exponent appearing in eq.~(\ref{eq:propagator_in_complex_path-integral}) takes the form $\tilde{R}/\hbar_{{\rm eff}}$ with $\tilde{R}={\cal O}(1)$. Following this recipe, eq.~(\ref{eq:propagator_in_complex_path-integral}) leads to
\begin{equation}     
\hbar_{{\rm eff}}=N^{-1},
\end{equation}
showing that in the present approach {\it the classical limit corresponds to the limit of large number of particles}. In the following, we complete the stationary analysis of the exact propagator valid in this $N \gg 1$ limit.

In eq.~(\ref{eq:propagator_in_complex_path-integral}) all integrals, that can and should be carried out exactly, are already performed, except for the integration over the initial phase of the first occupied single particle state. This integration has to be done exactly because of the $U(1)$ gauge symmetry, \textit{i.e.~}the freedom to multiply the wave function by an arbitrary global phase. In order to perform this integration, one first has to substitute the integrations over the real and imaginary part of $\phi_j^{(m)}$ by those over its modulus squared $J_j^{(m)}$ and phase $\varphi_j^{(m)}$ and then has to substitute the latter by $\theta_j^{(m)}-\theta_{j_1}^{(0)}$, where $j_1$ denotes the first initially occupied single particle state,
\begin{equation}
j_1=\min\left\{j\in\{1,2,\ldots\}:n^{(i)}_j=1\right\}.
\label{eq:definition_j1}
\end{equation}
These substitutions can be summarized as
\begin{equation}
\phi_j^{(m)}=\sqrt{J_j^{(m)}}\exp\left[{\rm i}\left(\theta_j^{(m)}-\theta_{j_1}^{(0)}\right)\right],
\label{eq:replacement_complex_amplitude_phase}
\end{equation}
for all $j$ and $m\geq1$, while for $m=0$,
\begin{align}
\phi_j^{(0)}&=n_j^{(i)}\exp\left[{\rm i}\left(\theta_j^{(0)}-\theta_{j_1}^{(0)}\right)\right] & \text{if } j\neq j_1, 
\label{eq:replacement_phase_initial} \\
\phi_{j_1}^{(0)}&=\exp\left({\rm i}\theta_{j_1}^{(0)}\right).
\label{eq:phi_initial_j1}
\end{align}
After these substitutions it is easy to see that the remaining dependence of the path integral on the global phase $\theta_{j_1}^{(0)}$ is given by $\exp[{\rm i}(N_f-N_i)\theta_{j_1}^{(0)}]$, with $N_{i/f}=\sum_jn_j^{(i/f)}$ being the initial, respectively, final total number of particles. Therefore, the integration over the global phase simply yields a factor $2\pi\delta_{N_f,N_i}$, which accounts for the conservation of the total particle number. The remaining integrals over $J_j^{(m)}$ and $\theta_j^{(m)}$ are then performed in stationary phase approximation, where (similar to the derivation of Stirling's approximation) for consistency and in order to include the behavior of the integrand especially for small occupations correctly, it is important to include the factors
\begin{equation}
\sqrt{J_j^{(m)}}=\exp\left[\log\left(J_j^{(m)}\right)/2\right]
\label{eq:inclusion_prefactors}
\end{equation}
in the stationarity analysis. For intermediate times, $1\leq m<M$, the stationarity conditions for $J_j^{(m)}$ and $\theta_j^{(m)}$ can be combined to the conditions
\begin{align}
{\rm i}\hbar\left(\phi_{j}^{(m)}-\phi_j^{(m-1)}\right)=&\tau\frac{\partial H_{cl}\left({\gV{\phi}^{(m)}}^\ast,\gV{\phi}^{(m-1)}\right)}{\partial{\phi_j^{(m)}}^\ast},
\label{eq:eom_discrete_intermediate} \\
-{\rm i}\hbar\left({\phi_{j}^{(m+1)}}^\ast-{\phi_j^{(m)}}^\ast\right)=&\tau\frac{\partial H_{cl}\left({\gV{\phi}^{(m+1)}}^\ast,\gV{\phi}^{(m)}\right)}{\partial{\phi_j^{(m)}}}.
\label{eq:eom_discrete_intermediate_conj}
\end{align}
In the same way, the conditions for $m=M$ can be written in the form of eq.~(\ref{eq:eom_discrete_intermediate}) with $m=M$ as well as the boundary condition
\begin{equation}
J_j^{(M)}=n_j^{(f)}.
\label{eq:bcf_discrete}
\end{equation}
Note that a linear combination of the stationarity conditions for $\theta_j^{(M)}$ and $J_j^{(M)}$ is required to get the stationary phase conditions in this form.

Since the integration over the initial phase is performed only for occupied states, and the amplitude of $\phi_j^{(0)}$ is equal to the initial occupation of the site $n_j^{(i)}$, the stationarity condition for $\theta_j^{(0)}$ yields eq.~(\ref{eq:eom_discrete_intermediate_conj}) with $m=0$. When finally taking the continuous limit $\tau\to0$, these conditions result in the equations of motion
\begin{align}
{\rm i}\hbar\dot{\gV{\phi}}(t)&=\frac{\partial H_{cl}\left(\gV{\phi}^\ast(t),\gV{\phi}(t)\right)}{\partial\gV{\phi}^\ast(t)},
\label{eq:eom} \\
-{\rm i}\hbar\dot{\gV{\phi}}^\ast(t)&=\frac{\partial H_{cl}\left(\gV{\phi}^\ast(t),\gV{\phi}(t)\right)}{\partial\gV{\phi}(t)},
\label{eq:eom_conj}
\end{align}
along with the boundary conditions
\begin{align}
\left|\phi_j(0)\right|^2&=n_j^{(i)}, & \left|\phi_j(t_f)\right|^2&=n_j^{(f)}
\label{eq:bc}
\end{align}
with $\phi_{j_1}(0)=1$. It is important to note that the equations of motion (\ref{eq:eom}) and (\ref{eq:eom_conj}) are complex conjugates of each other, such that for $J$ single particle states we get $J$ complex (or correspondingly $2J$ real) equations of motion with $2J$ real boundary conditions. Therefore one can always find at least one solution without the complexification necessary for the bosonic coherent state propagator \cite{Aguiar+Barangar,Garg+Braun}. Therefore, the classical Hamiltonian and action will also be real.

We also point out the key difference in the role of the boundary conditions in eq.~(\ref{eq:bc}) when compared with the derivation of the classical limit from the path integral in the standard first-quantized case. In the later, {\it boundary conditions are imposed at the level of the path integral} and therefore are not subject to the stationary phase conditions. Contrary to the bosonic case where this observation remains true \cite{cbs_fock}, here again we encounter that the classical limit of fermionic fields displays counter-intuitive features: the boundary conditions (\ref{eq:bc}) that allow for multiple solutions of (\ref{eq:eom}, \ref{eq:eom_conj}) are themselves obtained from a stationary phase argument, and the corresponding quantum fluctuations must be considered at the same footing as the fluctuations around the classical solutions.

Evaluating the exponent of the path integral along the stationary point (including all additional phase factors originating from the boundary terms $m=1,M$) then yields the classical action
\begin{align}
R_{\gamma}&\left(\V{n}^{(f)},\V{n}^{(i)};t_f\right)=\int\limits_{0}^{t_f}{\rm d}t\left[\hbar\gV{\theta}(t)\cdot\dot{\V{J}}(t)-H_{cl}\left(\gV{\phi}^\ast(t),\gV{\phi}(t)\right)\right],
\label{eq:action}
\end{align}
of the mean field trajectories defined by the equations of motion (\ref{eq:eom}) and the boundary conditions (\ref{eq:bc}). In eq.~(\ref{eq:action}) the real functions $\gV{\theta}(t)$ and $\V{J}(t)$ are defined through
\begin{equation}
\phi_j(t)=\sqrt{J_j(t)}\exp\left({\rm i}\theta_j(t)\right).
\label{eq:definition_J_theta_timedependent}
\end{equation}
It is worth to note, that the equations of motion (\ref{eq:eom},\ref{eq:eom_conj}) in these variables can also be written as the real equations
\begin{align}
\dot{\V{J}}(t)&=\frac{2}{\hbar}\frac{\partial H_{cl}\left(\gV{\phi}^\ast(t),\gV{\phi}(t)\right)}{\partial\gV{\theta}(t)},
\label{eq:eom_J} \\
\dot{\gV{\theta}}(t)&=-\frac{2}{\hbar}\frac{\partial H_{cl}\left(\gV{\phi}^\ast(t),\gV{\phi}(t)\right)}{\partial\V{J}(t)},
\label{eq:eom_theta}
\end{align}
where $\phi_j^\ast(t)$ and $\phi_j(t)$ should be understood as functions of $J_j(t)$ and $\theta_j(t)$ according to eq.~(\ref{eq:definition_J_theta_timedependent}). Thus, the classical trajectory lives on a symplectic manifold in phase space, which is here defined as $\{(\V{J},\gV{\theta}):J_{j=1,2,\ldots}\in[0,\infty),\theta_{j=1,2,\ldots}\in[0,2\pi)\}$. Moreover, the theory of canonical transformations \cite{Tabor} can be applied to show that the Poincar\'e-Cartan 1-form
\begin{equation}
\gV{\theta}\cdot{\rm d}\V{J}-H{\rm d}t
\label{eq:poincare-cartan}
\end{equation}
is invariant under canonical transformations.

The derivatives of the action can be found by applying the equations of motion to the integrand to read
\begin{align}
\frac{\partial R_{\gamma}\left(\V{n}^{(f)},\V{n}^{(i)};t_f\right)}{\partial\V{n}^{(i)}}&=-\hbar\gV{\theta}(0),
\label{eq:deriv_action_ni} \\
\frac{\partial R_{\gamma}\left(\V{n}^{(f)},\V{n}^{(i)};t_f\right)}{\partial\V{n}^{(f)}}&=\hbar\gV{\theta}(t_f),
\label{eq:deriv_action_nf} \\
\frac{\partial R_{\gamma}\left(\V{n}^{(f)},\V{n}^{(i)};t_f\right)}{\partial t_f}&=-E_\gamma,
\label{eq:deriv_action_t}
\end{align}
where $E_\gamma=H_{cl}\left(\gV{\phi}^\ast(0),\gV{\phi}(0)\right)$ is the energy of the trajectory.

Finally, the propagator eq.~(\ref{eq:definition_propagator_fock}) reads
\begin{equation}
K^{\rm sc}\left(\V{n}^{(f)},\V{n}^{(i)};t_f\right)=\sum\limits_{\gamma}^{}\mathcal{A}_{\gamma}\exp\left[\frac{\rm i}{\hbar}R_{\gamma}\left(\V{n}^{(f)},\V{n}^{(i)};t_f\right)\right],
\label{eq:semiclassical_propagator_before_fluctuation-integration}
\end{equation}
where the sum runs over all ``classical paths'' $\gamma$ which satisfy the equations of motion (\ref{eq:eom}) and the boundary conditions (\ref{eq:bc}), while $\mathcal{A}_{\gamma}$ is given by the still pending integrations over the second variation of the paths.
As is shown in appendix \ref{app:semiclassical_amplitude}, $\mathcal{A}_{\gamma}$ can be written as
\begin{align}
\mathcal{A}_{\gamma}=&\frac{1}{\sqrt{2\pi}^{N-1}}\exp\left\{\frac{\rm i}{2\hbar}\int\limits_{0}^{t_f}{\rm d}t{\rm Tr}\left[\frac{\partial^2H_{cl}}{\partial\gV{\phi}(t)^2}\mathbf{X}(t)\right]\right\} \nonumber \\
&\det\left\{\mathbf{I}_{N}+\exp\left[-2{\rm i}{\rm diag}\left(\mathbf{P}_{f}\gV{\theta}(t_f)\right)\right]\mathbf{P}_f\mathbf{X}(t_f)\mathbf{P}_f^{\rm T}\right\}^{-\frac{1}{2}},
\label{eq:semiclassical_prefactor_intermediately}
\end{align}
with $N=N_{i}=N_{f}$ being the total particle number and $\mathbf{I}_N$ the $N\times N$ unit matrix. Moreover, $\mathbf{P}_f$ is the matrix of the projector onto the subspace of the states which are occupied at final time, such that \textit{e.g.}
\begin{equation}
\mathbf{P}_f\V{n}^{(f)}=(\underbrace{1,\ldots,1}_{N})^{\rm T}.
\label{eq:example_for_action_of_Pf}
\end{equation}

For later reference, we also define $\mathbf{P}_i$, which is defined in the same way as $\mathbf{P}_f$, but selecting the initially occupied single particle states, as well as the complements $\bar{\mathbf{P}}_{i/f}$ of $\mathbf{P}_{i/f}$. With these matrices, one can also define the (orthonormal) matrix
\begin{equation}
\mathbf{Q}_{i/f}=\left(\begin{array}{c}
\bar{\mathbf{P}}_{i/f} \\
\mathbf{P}_{i/f}
\end{array}\right)
\label{eq:definition_Q},
\end{equation}
shifting all components of a vector corresponding to an initially (finally) unoccupied single particle state in front of all the others.

Finally in eq.~(\ref{eq:semiclassical_prefactor_intermediately}) $\mathbf{X}(t)$ satisfies the differential equation
\begin{align}
\dot{\mathbf{X}}(t)=&\frac{\rm i}{\hbar}\frac{\partial^2H_{cl}}{{\partial\gV{\phi}^\ast(t)}^2}-\frac{\rm i}{\hbar}\frac{\partial^2H_{cl}}{\partial\gV{\phi}^\ast(t)\partial\gV{\phi}(t)}\mathbf{X}(t) \nonumber \\
&-\frac{\rm i}{\hbar}\mathbf{X}(t)\frac{\partial^2H_{cl}}{\partial\gV{\phi}(t)\partial\gV{\phi}^\ast(t)}+\frac{\rm i}{\hbar}\mathbf{X}(t)\frac{\partial^2H_{cl}}{\partial\gV{\phi}(t)^2}\mathbf{X}(t),
\label{eq:deq_X}
\end{align}
with initial condition
\begin{equation}
\mathbf{X}(0)=\mathbf{Q}_i^{\rm T}
\left(\begin{array}{cc}
0 \\
 & \exp\left[2{\rm i}{\rm diag}\left(\mathbf{P}_i^\prime\gV{\theta}(0)\right)\right]
\end{array}\right)\mathbf{Q}_i.
\label{eq:ic_X}
\end{equation}
The same differential equation, however with different initial conditions, was encountered previously in derivations of a semiclassical propagator for bosonic many body systems in coherent state representation \cite{Aguiar+Barangar,Garg+Braun}. The solutions given there indicate, how to find $\mathbf{X}(t)$: Consider a solution $\boldsymbol\psi(t)$ of the equations of motion with initial conditions ${\bf Y}$ and ${\bf W}$, whereby each pair $(Y_j,W_j)$ are canonically conjugate variables. Possibilities for the choice of these pairs are \textit{e.g.}~$(\Re\psi_j(0),\Im\psi_j(0))$, where $\Re$ and $\Im$ denote the real and imaginary part, respectively, $(\left|\psi_j(0)\right|,\arg\psi_j(0))$ with $\arg\psi$ denoting the phase of $\psi$, or $(\psi_j(0),{\psi}^\ast_j(0))$. Then, the differential equation (\ref{eq:deq_X}) is solved by the function
\begin{equation}
-\frac{\partial{\boldsymbol\psi}(t)}{\partial\bf W}\left(\frac{\partial{\boldsymbol\psi}^\ast(t)}{\partial\bf W}\right)^{-1},
\label{eq:general_solution_X}
\end{equation}
evaluated at the initial conditions corresponding to the trajectory $\gamma$.

Finally, in order to find the solution for $\mathbf{X}(t)$, the variables ${\bf Y}$ and ${\bf W}$ need to be chosen such that for $t=0$, eq.~(\ref{eq:general_solution_X}) also satisfies the initial condition (\ref{eq:ic_X}), which yields
\begin{equation}
(Y_j,W_j)=\left\{\begin{array}{lc}
({\psi}_j(0),{\psi}_j^\ast(0)),\quad & \text{if } n_j^{(i)}=0 \text{ or } j=j_1 \\
(n_{j}^{(i)},\theta_{j}), & \text{else}.
\end{array}\right.
\label{eq:choice_initial_conditions}
\end{equation}
Eventually, the semiclassical amplitude $\mathcal{A}_{\gamma}$ can be written as
\begin{align}
\mathcal{A}_{\gamma}=&\sqrt{\det\left[\frac{1}{2\pi{\rm i}\hbar}\frac{\partial^2R_{\gamma}}{\partial\left(\mathbf{P}_f^{\prime}\V{n}^{(f)}\right)\partial\left(\mathbf{P}_i^{\prime}\V{n}^{(i)}\right)}\right]} \nonumber \\
&\sqrt{\det\mathbf{Q}_f\mathbf{Q}_i}\exp\left(\frac{\rm i}{2\hbar}\int\limits_{0}^{t_f}{\rm d}t{\rm Tr}\frac{\partial^2H_{cl}}{\partial\gV{\phi}^\ast\partial\gV{\phi}}\right) \nonumber \\
&\exp\left(\frac{\rm i}{2}\sum\limits_{j:n_j^{(f)}=1}^{}\theta_j(t_f)-\frac{\rm i}{2}\sum\limits_{j:n_j^{(i)}=1}^{}\theta_j(0)\right) \nonumber \\
&\det\left(\mathbf{A}-\mathbf{B}\mathbf{C}^{-1}\mathbf{D}\right)^{-\frac{1}{2}}.
\label{eq:semiclassical_prefactor}
\end{align}
with $\mathbf{P}_i^\prime$ and $\mathbf{P}_f^\prime$ being the matrices resulting from $\mathbf{P}_i$ and $\mathbf{P}_f$, respectively, by removing the first line. The determinant consisting of the matrices
\begin{align}
\mathbf{A}&=\frac{\partial\left(\bar{\mathbf{P}}_f\gV{\phi}^\ast(t_f),J_{\min\left\{j\in\{1,2,\ldots\}:n_j^{(f)}=1\right\}}(t_f)\right)}{\partial\left(\bar{\mathbf{P}}_i^{\prime}\gV{\phi}^\ast(0)\right)},
\label{eq:firstderivmatrix} \\
\mathbf{B}&=\frac{\partial\left(\bar{\mathbf{P}}_f\gV{\phi}^\ast(t_f),J_{\min\left\{j\in\{1,2,\ldots\}:n_j^{(f)}=1\right\}}(t_f)\right)}{\partial\left(\mathbf{P}_i^{\prime}\gV{\theta}(0)\right)},
\label{eq:secondderivmatrix} \\
\mathbf{C}&=\frac{\partial\left(\mathbf{P}_f^{\prime}\V{J}(t_f)\right)}{\partial\left(\mathbf{P}_i^{\prime}\gV{\theta}(0)\right)},
\label{eq:thirdderivmatrix} \\
\mathbf{D}&=\frac{\partial\left(\mathbf{P}_f^{\prime}\V{J}(t_f)\right)}{\partial\left(\bar{\mathbf{P}}_i^{\prime}\gV{\phi}^\ast(0)\right)}.
\label{eq:fourthderivmatrix}
\end{align}
accounts for the vacuum fluctuations that have been treated exactly.
Note that in eq.~(\ref{eq:semiclassical_prefactor}) the Solari-Kochetov extra-phase
\begin{equation}
\exp\left(\frac{\rm i}{2\hbar}\int\limits_{0}^{t_f}{\rm d}t{\rm Tr}\frac{\partial^2H_{cl}}{\partial\gV{\phi}^\ast\partial\gV{\phi}}\right)
\end{equation}
typically arises in a semiclassical approximation of the propagator in coherent state representation \cite{Aguiar+Barangar,Solari,Kochetov,Vieira+Sacramento}, while in the standard (first quantized) van-Vleck-Gutzwiller propagator \cite{Gut1970}, this phase is absent, due to the Weyl (symmetric) ordering of the Hamiltonian with respect to position and momentum operators. For Bosons, the Solari-Kochetov phase can be absorbed in the action by replacing the bosoinc creation and annihilation operators according to $\hat{a}_j^\dagger\hat{a}_{j^\prime}^{}\to(\hat{a}_j^\dagger\hat{a}_{j^\prime}^{}+\hat{a}_{j^\prime}^{}\hat{a}_j^\dagger)/2$ \cite{Aguiar+Barangar}, which corresponds to Weyl ordering of the quantum Hamiltonian. In the same way, for the propagator in spin coherent states, this phase is absent in Weyl ordering \cite{Pletyukhov_extra-phase}. However, this vanishing of the Solari-Kochetov phase in these cases is due to the fact that the classical Hamiltonian is obtained out of the quantum one by the simple replacements $\hat{a}_j^\dagger\to\phi_j^\ast$ and $\hat{a}_j^{}\to\phi_j^{}$, which is not valid here. Therefore, it seems that here this phase can not be eliminated by changing the chosen ordering of the fermionic creation and annihilation operators.

Due to their definition eq.~(\ref{eq:definition_Q}), the determinants $\det\mathbf{Q}_{i/f}$ depend only on the choice of the initial and final occupations and accept only the values $\pm1$. Note that this sign also depends on the definition of the Fock states, while the product of both depends only on the relative changes between the initial and final state and therefore is independent of the exact choice of ordering of the single particle states.

It is important to notice that in eq.~(\ref{eq:semiclassical_prefactor}) the determinant $\det(\mathbf{A}-\mathbf{B}\mathbf{C}^{-1}\mathbf{D})$ depends only on the derivatives of the values of the trajectory at final time with respect to the initial conditions and should therefore be possible to calculate in an actual application. Moreover, we expect that this determinant is just the product of the exponentials of the final and initial phases of the final unoccupied states, which can be set to zero. Thus, we assume this determinant to be equal to one. However, up to now we did not succeed in proofing this conjecture rigorously and therefore, we will keep this determinant in the following.

%% file: transition_probability.tex
\section{Transition probability}
\label{sec:transition_prob}
\subsection{General semiclassical treatment}
Knowing the propagator enables us, in principle, to calculate the quantum probability to measure the Fock state $\V{n}^{(f)}$ after preparing the system of spin-$1/2$ particles in the initial Fock state $\V{n}^{(i)}$ and letting it evolve for some time $t$. Computing this probability is usually non-trivial, since the single particle states can on the one hand be chosen arbitrarily, and may thus not necessarily be eigenstates of the single-particle Hamiltonian and on the other hand interactions in general induce a coupling between different single particle states. This probability is given by the modulus square of the overlap between the time evolved state and $\ket{\V{n}^{(f)}}$,
\begin{equation}
P\left(\V{n}^{(f)},\V{n}^{(i)};t_f\right)=\left|\Braket{\V{n}^{(f)}|\hat{K}\left(t_f\right)|\V{n}^{(i)}}\right|^2.
\label{eq:definition_transition_probability}
\end{equation}
Using the semiclassical approximation (\ref{eq:semiclassical_propagator_before_fluctuation-integration}), it is given by a double sum over trajectories,
\begin{equation}
P\left(\V{n}^{(f)},\V{n}^{(i)};t_f\right)\approx\sum\limits_{\gamma,\gamma^\prime}^{}\mathcal{A}_{\gamma^{}}\mathcal{A}_{\gamma^\prime}^\ast\exp\left[\frac{\rm i}{\hbar}\left(R_{\gamma^{}}-R_{\gamma^\prime}\right)\right].
\label{eq:transition_probability_semiclassically}
\end{equation}
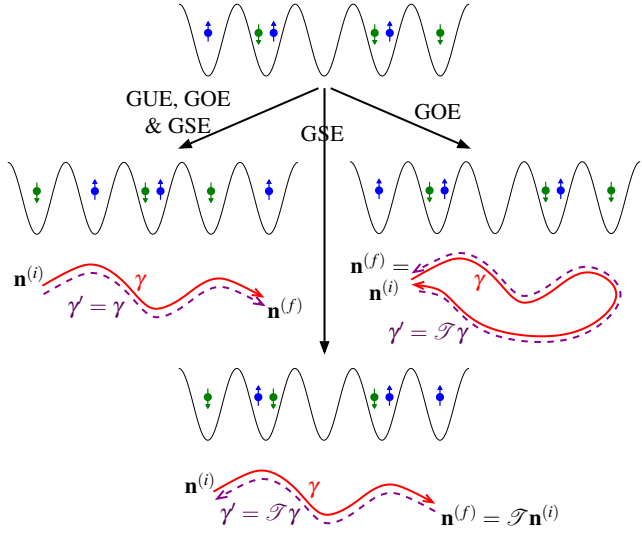
\begin{figure}
\includegraphics[width=0.48\textwidth]{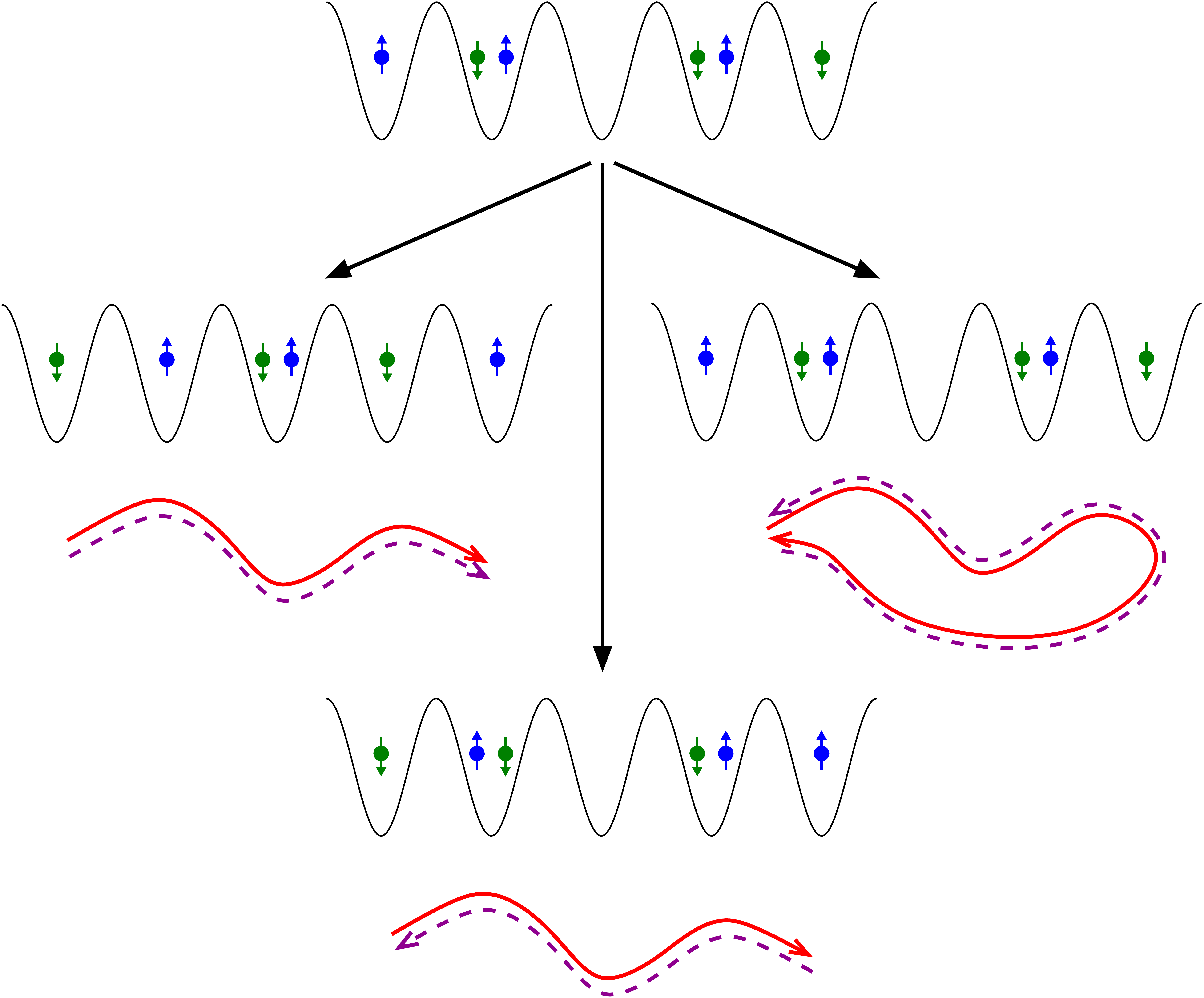}
\put(-235,90){${\bf n}^{(i)}$}
\put(-140,78){${\bf n}^{(f)}$}
\put(-109,95){${\bf n}^{(f)}=$}
\put(-102,85){${\bf n}^{(i)}$}
\put(-171,12){${\bf n}^{(i)}$}
\put(-75,0){${\bf n}^{(f)}=\mathcal{T}{\bf n}^{(i)}$}
\put(-193,158){GUE, GOE}
\put(-186,148){\& GSE}
\put(-85,153){GOE}
\put(-127.5,145){GSE}
\put(-190,90){$\textcolor{red}{\gamma}$}
\put(-215,80){$\textcolor[rgb]{0.35,0,0.35}{\gamma^\prime=\gamma}$}
\put(-63,90){$\textcolor{red}{\gamma}$}
\put(-95,70){$\textcolor[rgb]{0.35,0,0.35}{\gamma^\prime=\mathcal{T}\gamma}$}
\put(-125,12){$\textcolor{red}{\gamma}$}
\put(-158,2){$\textcolor[rgb]{0.35,0,0.35}{\gamma^\prime=\mathcal{T}\gamma}$}
\caption{The quantum transition in a system of spin-$1/2$ particles in the semiclassical limit. A trajectory $\gamma$ is paired with a partner trajectory $\gamma^\prime$, where $\gamma^\prime$ can be either $\gamma$ itself, or its time reverse. The annotations at the arrows indicate the symmetry class required for the corresponding pairing to be present.}
\label{fig:pairing}
\end{figure}
Upon applying an energy or disorder average, the action difference gives rise to huge oscillations, such that most contributions to the averaged double sum will cancel, except if the paths $\gamma$ and $\gamma^\prime$ are correlated. The types of trajectory pairs, which we will consider in the following are depicted in fig.~\ref{fig:pairing}. The simplest type of correlation arises for $\gamma=\gamma^\prime$. This is known as the diagonal approximation \cite{da}. The second derivatives of the action with respect to the initial and final Fock state in the prefactor can then be used to transform the sum over trajectories into an integration over the initial phases. Then the diagonal approximation yields,
\begin{align}
P&_{cl}\left(\V{n}^{(f)},\V{n}^{(i)};t_f\right)= \nonumber \\
&\int\limits_{0}^{2\pi}{\rm d}^{N-1}\theta^{(i)}\det\left(\mathbf{A}-\mathbf{B}\mathbf{C}^{-1}\mathbf{D}\right)^{-1}\delta\left(\left|\gV{\phi}(t_f)\right|^2-\V{n}^{(f)}\right),
\label{eq:classical_probability}
\end{align}
which we will refer to as classical probability. Here $\gV{\phi}(t)$ is the solution of the equations of motion eq.~(\ref{eq:eom}) with the initial condition $\phi_j(0)=\sqrt{n_j^{(i)}}\exp\left({\rm i}\theta_j^{(i)}\right)$. It is worth to notice that the exact treatment of the vacuum fluctuations gives rise to a renormalization of the transition probability by the additional factor $\det\left(\mathbf{A}-\mathbf{B}\mathbf{C}^{-1}\mathbf{D}\right)^{-1}$.

Further pairs of correlated trajectories are those given by $\gamma$ and its time reverse, $\gamma^\prime=\mathcal{T}\gamma$. However, the time reverse of a trajectory exists only if the system is time reversal symmetric. Moreover, the initial and final occupations, respectively, of both trajectories in the double sum of eq.~(\ref{eq:transition_probability_semiclassically}) have to be the same. On the other hand, if $\gamma$ has initial occupations $\V{n}^{(i)}$ and final occupations $\V{n}^{(f)}$, the initial occupations of its time reverse are given by the time reverse of $\V{n}^{(f)}$ and the final ones by the time reverse of $\V{n}^{(i)}$. Therefore, in order to pair $\gamma$ with its time reverse, we need time reversal symmetry and also the final Fock state has to be the time reverse of the initial one. To this end, one has to replace the sum over trajectories from $\V{n}^{(i)}$ to $\V{n}^{(f)}$ by a sum over trajectories ending at the Fock state $\mathcal{T}\V{n}^{(i)}$ originating from time reversing the initial one. To this end, the actions in the exponential need to be expanded in the final Fock state around $\mathcal{T}\V{n}^{(i)}$ up to linear order, while the prefactor is assumed to vary only very slightly with $\V{n}^{(f)}$, such that it can be simply replaced by $\mathcal{T}\V{n}^{(i)}$. For pairs $\gamma^\prime=\mathcal{T}\gamma$ this procedure then gives the contribution
\begin{align}
\sum\limits_{\gamma}^{}&\mathcal{A}_{\gamma^{}}\mathcal{A}_{\mathcal{T}\gamma^{}}^\ast\exp\left(\frac{\rm i}{\hbar}\Delta R_{}\right)\times \nonumber \\
&\times\exp\left[{\rm i}\left(\gV{\theta}^{(\gamma)}(t_f)-\gV{\theta}^{(\mathcal{T}\gamma)}(t_f)\right)\cdot\left(\V{n}^{(f)}-\mathcal{T}\V{n}^{(i)}\right)\right],
\label{eq:cbs_contribution_before_average}
\end{align}
with the difference $\Delta R=R_\gamma-R_{\mathcal{T}\gamma}$ in the actions of $\gamma$ and $\mathcal{T}\gamma$. Since for time reversal symmetric systems, the energy of a trajectory and its time-reverse is the same, we easily get
\begin{equation}
\Delta R=\hbar\int\limits_{0}^{t_f}{\rm d}t\left(\gV{\theta}^{(\gamma)}\cdot\dot{\V{J}}^{(\gamma)}-\gV{\theta}^{(\mathcal{T}\gamma)}\cdot\dot{\V{J}}^{(\mathcal{T}\gamma)}\right).
\label{eq:action_difference_time-reverse_general}
\end{equation}
In the next steps, we assume -- in accordance with the cases considered below -- that the difference $\Delta R$, is independent of the trajectory. This is usually the case, since the second part of the integral in the action difference can be related with the first one by making use of the nature of the time reversal operation. However, as we will see later, $\Delta R$ does not vanish in general. Moreover, we can savely assume that $\gV{\theta}^{(\mathcal{T}\gamma)}(t_f)$ depends on the initial phases of $\gamma$, only (and through them on the initial Fock state).

Upon disorder average, the phases $\theta_j^{(\gamma)}(t_f)$ behave, for chaotic systems, like linearly distributed random variables between $0$ and $2\pi$. Thus, treating them as random variables and performing the average, yields a $\delta_{\V{n}^{(f)},\mathcal{T}\V{n}^{(i)}}$, such that one gets after utilizing the second derivative of the action again
\begin{equation}
P_{cl}\left(\V{n}^{(f)},\V{n}^{(i)};t_f\right)\delta_{\V{n}^{(f)},\mathcal{T}\V{n}^{(i)}}\exp\left(\frac{\rm i}{\hbar}\Delta R\right).
\label{eq:cbs_contribution_TR_not_specified}
\end{equation}
The action difference $\Delta R$ strongly depends on whether the system is diagonal in spin space or not.
\subsection{Systems diagonal in spin space}
\label{subsec:diagonal_spin-space}
If the system is diagonal in spin space, \textit{i.e.}~the Hamiltonian does not consist of terms giving rise to spin-flips, the time reversal operation amounts to a complex conjugation only, and therefore
\begin{equation}
\mathcal{T}\V{n}^{(i)}=\V{n}^{(i)}.
\label{eq:time-reversal_initial_fock-state_diagonal_spin-space}
\end{equation}
It also implies that, on the classical level, the time reverse of $\gV{\phi}(t)$ is given by $\gV{\phi}^\ast(t_f-t)$. With this information, it is easy to prove that time reversed paths have the same action, $\Delta R=0$. Thus in semiclassical approximation the averaged transition probability for a spin-diagonal system is given by
\begin{equation}
P\left(\V{n}^{(f)},\V{n}^{(i)};t_f\right)\approx P_{cl}\left(\V{n}^{(f)},\V{n}^{(i)};t_f\right)\left(1+\delta_{\V{n}^{(f)},\V{n}^{(i)}}\right).
\label{eq:transition_probabilit_spin-diagonal}
\end{equation}
This is, apart from apart from the renormalization of the classical transition probability due to the exact treatment of the vacuum (see eq.~(\ref{eq:classical_probability})), exactly the same result found previously for bosonic, spinless systems \cite{cbs_fock}.
\subsection{Systems non-diagonal in spin space}
\label{non-diagonal_spin-space}
If the system's Hamiltonian is non-diagonal in spin space, the time reversal operation is not just complex conjugation, but also demands an exchange of the spin-up and spin-down components while at the same time introducing a relative minus sign between them,
\begin{equation}
\hat{T}=\left[\prod\limits_{j}\left(-{\rm i}\hat{\sigma}_{j,y}\right)\right]\hat{K}.
\label{eq:tr_non-diagonal_spin-space}
\end{equation}
Here $\hat{\sigma}_{j,y}$ is the $y$-Pauli matrix for the $j$-th state and $\hat{K}$ denotes complex conjugation. Important examples of systems with such time reversal operations are for instance systems with a Rashba spin-orbit coupling \cite{Rashba}, whcih is of key importance in semiconductor spintronics, but more recently has also been realized using ultra-cold atoms \cite{Spielman}.

On the classical level, this means that the time reversal of $\gV{\phi}=(\gV{\phi}_{\uparrow},\gV{\phi}_{\downarrow})^{\rm T}$, where $\gV{\phi}_{\uparrow(\downarrow)}$ is the vector containing all spin-up (spin-down) components of $\gV{\phi}$, is given by
\begin{equation}
\mathcal{T}\left(\begin{array}{c}
\gV{\phi}_{\uparrow}(t) \\
\gV{\phi}_{\downarrow}(t)
\end{array}\right)=\left(\begin{array}{c}
-\gV{\phi}_{\downarrow}^\ast(t_f-t) \\
\gV{\phi}_{\uparrow}^\ast(t_f-t)
\end{array}\right),
\label{eq:tr-trajectory_non-diagonal_spin-space}
\end{equation}
and therefore also
\begin{equation}
\mathcal{T}\V{n}^{(i)}=\mathcal{T}\left(\begin{array}{c}
\V{n}^{(i)}_{\uparrow} \\
\V{n}^{(i)}_{\downarrow}
\end{array}\right)=\left(\begin{array}{c}
\V{n}^{(i)}_{\downarrow} \\
\V{n}^{(i)}_{\uparrow}
\end{array}\right).
\label{eq:timereverse_ni}
\end{equation}
For the action difference, this yields
\begin{equation}
\Delta R=\pi\hbar\sum\limits_{j}^{}\left[\left(\mathcal{T}\V{n}^{(i)}\right)_{j,\uparrow}-n^{(i)}_{j,\uparrow}\right]=\pi\hbar\left(N_{\downarrow}-N_{\uparrow}\right),
\label{eq:action_difference_non-diagonal_spin-space}
\end{equation}
where $N_{\uparrow(\downarrow)}$ is the total number of spin-up (spin-down) particles in the initial state.

Thus, invoking the widely used nomenclature of the random matrix symmetry classes and quantum chaos \cite{Haake}, one finally finds for the averaged transition probability in semiclassical approximation
\begin{align}
P&\left(\V{n}^{(f)},\V{n}^{(i)};t_f\right)= \nonumber \\
&P^{(cl)}\left(\V{n}^{(f)},\V{n}^{(i)};t_f\right)\left\{\begin{array}{ll}
1 & \text{, GUE} \\
\left[1+\delta_{\V{n}^{(f)},\V{n}^{(i)}}\right] & \text{, GOE} \\
\left[1+(-1)^N\delta_{\V{n}^{(f)}_{\downarrow},\V{n}^{(i)}_{\uparrow}}\delta_{\V{n}^{(f)}_{\downarrow},\V{n}^{(i)}_{\uparrow}}\right] & \text{, GSE}.
\end{array}\right.
\label{eq:transition_probability_semiclassical}
\end{align}
Here GUE (Gaussian Unitary Ensemble) means that the average runs over systems without time reversal symmetry, while for GOE (Gaussian Orthogonal Ensemble) and GSE (Gaussian Symplectic Ensemble) the average is over time reversal invariant spin-$1/2$ systems, which are diagonal and non-diagonal in spin space, respectively. This result and in particular the origin of the deltas is illustrated in fig.~\ref{fig:pairing}.

It is important to note that, the probability to find $\V{n}^{(f)}=\mathcal{T}\V{n}^{(i)}$ is zero on average for the GSE case, if $N$ is odd. However, the transition probability is a strictly positive quantity. Therefore, in order to become zero on average, it has to be zero for each disorder realization. In other words, for a time reversal symmetric system, which is non-diagonal in spin space, the transition from an initial Fock state to its spin reversed version is semiclassically prohibited,
\begin{equation}
\Braket{\hat{T}\V{n}^{(i)}|\hat{K}\left(t_f\right)|\V{n}^{(i)}}=0
\label{eq:prohibited_transition}
\end{equation}
for an odd total number of particles. This is consistent with
\begin{equation}
\Braket{\hat{T}\V{n}|\hat{H}|\V{n}}=0.
\label{eq:kramers_degeneracy}
\end{equation}
Similar to the proof of Kramer's degeneracy \cite{Sakurai}, one can show that eq.~(\ref{eq:kramers_degeneracy}) implies that for an odd number of particles and a symplectic time reversal symmetry, the transition from a Fock state to its spin reversed version is exactly forbidden quantum mechanically.

On the other hand, if the total number of particles is even, and hence the total spin is integer, the transition probability is always enhanced by a factor of two compared to the classical one, if the final Fock state is the time reversed version of the initial one.

%% file: conclusion.tex
\section{Conclusions}
\label{sec:conclusion}
We presented a rigorous derivation of fermionic path integrals representing quantum transition amplitudes in Fock space in terms of unrestricted, commuting complex fields. In the context of semiclassical approaches we believe that this result represents an important improvement over previous approaches. First, we replace the anticommuting (Grassmann) variables, usually assumed to be the most natural representation of a fermionic path integral, by complex variables in the path integral. In this way, the propagator can be given a direct physical interpretation as a complex-valued amplitude. Second, the path integral is unrestricted (defined over the whole complex plane) and therefore avoids the complications due to the definition of path integrals in compact phase spaces.

Most notably, in the approach presented here a Hamiltonian classical limit can be identified which leads to real actions and therefore explicit interference. After applying the stationary phase approximations to the path integral. In the semiclassical limit (of large particle number), we are able to derive as our major result a van Vleck-Gutzwiller type propagator for fermionic quantum fields.  

In contrast to the approaches of \cite{miller_classical,miller_semiclassical}, here the semiclassical approximation as well as the classical limit are obtained from an exact path integral. However, there is still a freedom of choice for the classical Hamiltonian, which should be investigated further. Hence we do not exclude the possibility, that by a certain choice, the classical limits of \cite{miller_classical,miller_semiclassical} can be recovered. Moreover it remains to be explored, which classical limit is best suited for calculations and simulations. This may actually even depend on the actual problem at hand.

In Sec.~\ref{sec:transition_prob} we applied our results to the calculation of transition probabilities in the fermionic Fock space, and found a rich dependence of many-body interference effects on the universality class of the system. For systems with spin-orbit interaction that belong to the symplectic class, our results predict the exact cancellation of the transition probability between time-reversed many-body states, if the total number of particles is odd. This prediction that can be independently demonstrated to be a consequence of Kramer's degeneracy, is a very stringent test for the correctness of our approach. If the total particle number is even, however, the same transition is not only allowed, but its probability is enhanced by a factor of two compared to the transitions to other states. For systems without spin-flip mechanisms, we recover the coherent backscattering previously found for bosons \cite{cbs_fock}. Upon destroying time reversal symmetry all this effects vanish, and the transition probability profile can be assumed to be more or less constant for all Fock states.

Finally, we would like to note that, although the path integral eq.~(\ref{eq:propagator_in_complex_path-integral}) is restricted to the particle picture, \textit{i.e.}~to the case that a particle is defined through an occupied single particle state, it is also possible to construct a path integral in the hole picture (for more details see appendix \ref{subapp:hole-picture}), where a particle is defined as an unoccupied single particle state.

The {\it major} principle restriction of applicability of our approach is that the number of fermions $N \gg 1$ should be large enough (our experience in the bosonic case indicates that $N \sim 10$ is enough). Therefore, within this regime, electronic systems such as quantum dots, coupled discrete systems like spin chains modeled by Heisenberg or Ising type Hamiltonians, and molecular systems described by a discrete set of single-particle orbitals can be addressed. Still, then exist practical limitations of semiclassical approaches in concrete applications, related, \textit{e.g.}, to the solution of the shooting problem and the correct evaluation of amplitudes and Maslov indexes. We hope that our approach is still beneficial for the Chemical Physics community.

Finally, we remark that for treating emergent universal quantum fluctuations in mesoscopic systems we only need to verify that the classical limit displays chaotic behavior, a substantially easier task.

Further applications of the semiclassical methods along the lines presented here like the description of many-body spin echoes \cite{TomS} are presently under investigation.

%% file: appendix.tex
\newcommand{\rmi}{{\rm i}}
\newcommand{\rme}{{\rm e}}
\newcommand{\rmd}{{\rm d}}
\newcommand{\rbr}[1]{\left(#1\right)}
\newcommand{\cbr}[1]{\left[#1\right]}
\newcommand{\tbr}[1]{\left\{#1\right\}}
\newcommand{\abs}[1]{\left|#1\right|}
\newcommand{\olap}[2]{\Braket{#1|#2}}
\newcommand{\nfv}{{\bf n}^{(f)}}
\newcommand{\nfj}[1]{n^{(f)}_{#1}}
\newcommand{\niv}{{\bf n}^{(i)}}
\newcommand{\nij}[1]{n^{(i)}_{#1}}
\newcommand{\chikv}[1]{{\boldsymbol\chi}^{(#1)}}
\newcommand{\zetakv}[1]{{\boldsymbol\zeta}^{(#1)}}
\newcommand{\phikv}[1]{{\boldsymbol\phi}^{(#1)}}
\newcommand{\mukv}[1]{{\boldsymbol\mu}^{(#1)}}
\newcommand{\thetakv}[1]{{\boldsymbol \theta}^{(#1)}}
\newcommand{\Jkv}[1]{{\bf J}^{(#1)}}
\newcommand{\chikj}[2]{\chi^{(#1)}_{#2}}
\newcommand{\zetakj}[2]{\zeta^{(#1)}_{#2}}
\newcommand{\phikj}[2]{\phi^{(#1)}_{#2}}
\newcommand{\mukj}[2]{\mu^{(#1)}_{#2}}
\newcommand{\thetakj}[2]{\theta^{(#1)}_{#2}}
\newcommand{\Jkj}[2]{J^{(#1)}_{#2}}
\newcommand{\timestep}{\tau}
\newcommand{\pisteps}{M}
\newcommand{\pilimit}{\lim\limits_{\pisteps\to\infty}}
\newcommand{\Sites}{J}
\newcommand{\sites}{j}
\newcommand{\picount}{m}
\newcommand{\prodl}[2]{\prod\limits_{#1}^{#2}}
\newcommand{\sul}[2]{\sum\limits_{#1}^{#2}}
\newcommand{\intg}[3]{\int\limits_{#1}^{#2}\!\rmd#3}
\newcommand{\conjg}[1]{\left.#1\right.^\ast}
\newcommand{\pdiff}[2]{\frac{\partial#1}{\partial#2}}
\newcounter{subcounter}
\newcounter{storecounter}
\section{Derivation of the path integral}
\label{app:pathintegral}
For simplicity, in this section, we assume a quantum hamiltonian given by
\begin{equation}
\hat{H}=\sul{\alpha,\beta}{}h_{\alpha\beta}\hat{c}_{\alpha}^{\dagger}\hat{c}_{\beta}^{}+\sul{\substack{\alpha,\beta \\ \alpha\neq\beta}}{}U_{\alpha\beta}\hat{c}_{\alpha}^{\dagger}\hat{c}_{\beta}^{\dagger}\hat{c}_{\beta}^{}\hat{c}_{\alpha}^{}.
\label{eq:quantum_hamiltonian_app}
\end{equation}
The result for a non-diagonal interaction $U_{\alpha\beta\gamma\nu}$, however, is given in appendix \ref{app:classical_hamiltonians}
In order to get from eq.~(\ref{eq:propagator_in_grassmann_path-integral}) to the complex path integral eq.~(\ref{eq:propagator_in_complex_path-integral}), the following two integrals with $j,j^\prime\in\mathbb{N}_0$, will be inserted:
\begin{align}
&\intg{0}{2\pi}{\theta}\intg{}{}{^2\phi}\exp\rbr{-\abs{\phi}^2+\conjg{\phi}\rme^{\rmi\theta}-\rmi j\theta}\phi^{j^\prime}=2\pi^2\delta_{j,j^\prime}
\label{eq:inserted_integral1} \\
&\intg{}{}{^2\phi}\intg{}{}{^2\mu}\exp\rbr{-\abs{\phi}^2-\abs{\mu}^2+\conjg{\phi}\mu}\phi^j\rbr{\conjg{\mu}}^{j^\prime}=\pi^2j!\delta_{j,j^\prime},
\label{eq:inserted_integral2}
\end{align}
Thereby $\rmd^2\mu=\rmd\Re{\mu}\rmd\Im{\mu}$, \textit{i.e.}~the integrations over $\phi$ and, in the second case, over $\mu$ run over the whole complex plane. One should notice, that the first of these two integrals is just the second one, but with the modulus of $\mu$ already integrated out.

The first of these two integrals is used to decouple $\zetakv{0}$ from $\zetakv{1}$ by the following identity:

\begin{widetext}
\begin{align}
\int&\rmd^{2\Sites}\zetakj{0}{}\exp\rbr{-\conjg{\zetakv{0}}\cdot\zetakv{0}}\cbr{\prodl{\sites=1}{\Sites}\rbr{1+\conjg{\chikj{0}{\sites}}\zetakj{0}{\sites}}}\prodl{\sites=1}{\Sites}\rbr{\conjg{\zetakj{0}{\sites}}}^{\nij{\sites}}= \nonumber \\
&\int\frac{\rmd^{2N_i}\phikj{0}{}}{\pi^{N_i}}\int\limits_{0}^{2\pi}\frac{\rmd^{N_i}\thetakj{i}{}}{\left(2\pi\right)^{N_i}}\int\rmd^{2\Sites}\zetakj{0}{}\exp\rbr{-\conjg{\zetakv{0}}\cdot\zetakv{0}-\abs{\phikv{0}}^2+\conjg{\phikv{0}}\cdot\mukv{0}}\cbr{\prodl{\sites=1}{\Sites}\rbr{1+\conjg{\chikj{0}{\sites}}\phikj{0}{\sites}}}\cbr{\prodl{\sites=0}{\Sites-1}\rbr{1+\zetakj{0}{\Sites-\sites}\conjg{\mukj{0}{\Sites-\sites}}}}\prodl{\sites=1}{\Sites}\rbr{\conjg{\zetakj{0}{\sites}}}^{\nij{\sites}},
\label{eq:insert_integrals_0}
\end{align}
\end{widetext}
with $\mukj{0}{\sites}=\nij{\sites}\exp\rbr{\rmi\thetakj{i}{\sites}}$ for all $\sites\in\{1,\ldots,\Sites\}$, where $\Sites$ is the number of single particle states taken into account. Note that here, for the initially unoccupied single particle states, the phases $\thetakj{i}{\sites}$ are arbitrary but fixed, \textit{e.g.}~to zero, while the integration runs only over those initial phases $\thetakj{i}{\sites}$, for which $\nij{l}=1$. In this way, the integrals, that have to be performed exactly, in order to get a reasonable and correct semiclassical approximation for the propagator are already done, and do not have to be carried out later.

For the $N_i=\sum_{\sites=1}^{\Sites}\nij{\sites}$ initially occupied single particle states, the identity follows directly from eq.~(\ref{eq:inserted_integral1}), while for the unoccupied ones, it is important to notice, that the term $\conjg{\chikj{0}{\sites}}\zetakj{0}{\sites}$ does vanish when integrating over $\zetakv{0}$. This is because of the properties of the Grassmann integrals eq.~(\ref{eq:vanishing_grassmann-integrals}) and the fact, that there is no $\conjg{\zetakj{0}{\sites}}$ for those components, for which $\nij{\sites}=0$.

The thus obtained expression is the starting point for an iterative insertion of integrals of the form of eq.~(\ref{eq:inserted_integral2}). For $1\leq\picount<\pisteps$, an evaluation of the overlaps and matrix elements of eq.~(\ref{eq:propagator_in_grassmann_path-integral}) containing $\zetakv{\picount}$ yields the following expression:

\begin{widetext}
\begin{align}
&\cbr{\prodl{\sites=1}{\Sites}\rbr{1+\conjg{\chikj{\picount}{\sites}}\zetakj{\picount}{\sites}}}\cbr{1-\frac{\rmi\tau}{\hbar}\sul{\alpha,\beta=1}{\Sites}\left(h_{\alpha\beta}^{(\picount-1)}\conjg{\zetakj{\picount}{\alpha}}\chikj{\picount-1}{\beta}+U_{\alpha\beta}^{(\picount-1)}\conjg{\zetakj{\picount}{\alpha}}\conjg{\zetakj{\picount}{\beta}}\chikj{\picount-1}{\beta}\chikj{\picount-1}{\alpha}\right)}\prodl{\sites=1}{\Sites}\rbr{1+\conjg{\zetakj{\picount}{\sites}}\chikj{\picount-1}{\sites}}= \nonumber \\
&\qquad\left[a^{(\picount)}-\frac{\rmi\timestep}{\hbar}\sul{\alpha}{}h_{\alpha\alpha}^{(\picount-1)}b_\alpha^{(\picount)}-\frac{\rmi\timestep}{\hbar}\sul{\substack{\alpha,\beta \\ \alpha\neq\beta}}{}h_{\alpha\beta}^{(\picount-1)}c_{\alpha\beta}^{(\picount)}-\frac{\rmi\timestep}{\hbar}\sul{\substack{\alpha,\beta \\ \alpha\neq\beta}}{}U_{\alpha\beta}^{(\picount-1)}d_{\alpha\beta}^{(\picount)}\right]\prodl{\sites=1}{\Sites}\rbr{1+\conjg{\zetakj{\picount}{\sites}}\chikj{\picount-1}{\sites}}.
\label{eq:abbreviation_mth_factor}
\end{align}
With the help of the integral eq.~(\ref{eq:inserted_integral2}), the coefficients $a^{(\picount)}$, $b^{(\picount)}$, $c^{(\picount)}$ and $d^{(\picount)}$ can successively -- starting from $\picount=1$ -- be written as
\begin{align}
a^{(\picount)}=&\int\frac{\rmd^{2\Sites}\mukj{\picount}{}}{\pi^\Sites}\int\frac{\rmd^{2\Sites}\phikj{\picount}{}}{\pi^\Sites}\cbr{\prodl{\sites=1}{\Sites}\rbr{1+\conjg{\chikj{\picount}{\sites}}\phikj{\picount}{\sites}}}\exp\rbr{-\abs{\phikv{\picount}}^2-\abs{\mukv{\picount}}^2+\conjg{\phikv{\picount}}\cdot\mukv{\picount}}\prodl{\sites=0}{\Sites-1}\cbr{1+\zetakj{\picount}{\Sites-\sites}\sul{k=1}{\infty}\frac{1}{k!}\rbr{\conjg{\mukj{\picount}{\Sites-\sites}}}^{k}\rbr{\phikj{\picount-1}{\Sites-\sites}}^{k-1}},
\label{eq:insert_integral_am} \\
b_{\alpha}^{(\picount)}=&\int\frac{\rmd^{2\Sites}\mukj{\picount}{}}{\pi^\Sites}\int\frac{\rmd^{2\Sites}\phikj{\picount}{}}{\pi^\Sites}\exp\rbr{-\abs{\phikv{\picount}}^2-\abs{\mukv{\picount}}^2+\conjg{\phikv{\picount}}\cdot\mukv{\picount}}\conjg{\zetakj{\picount}{\alpha}}\chikj{\picount-1}{\alpha}\left\{\prodl{\sites=0}{L-\alpha-1}\cbr{1+\zetakj{\picount}{\Sites-\sites}\sul{k=1}{\infty}\frac{1}{k!}\rbr{\conjg{\mukj{\picount}{\Sites-\sites}}}^{k}\rbr{\phikj{\picount-1}{\Sites-\sites}}^{k-1}}\right\} \nonumber \\
&\quad\cbr{1+\zetakj{\picount}{\alpha}\sul{k=1}{\infty}c_k^{(1)}\rbr{\phikj{\picount-1}{\alpha}}\rbr{\conjg{\mukj{\picount}{\alpha}}}^{k}}\left\{\prodl{\sites=\Sites-\alpha+1}{\Sites-1}\cbr{1+\zetakj{\picount}{\Sites-\sites}\sul{k=1}{\infty}\frac{1}{k!}\rbr{\conjg{\mukj{\picount}{\Sites-\sites}}}^{k}\rbr{\phikj{\picount-1}{\Sites-\sites}}^{k-1}}\right\}\cbr{\prodl{\sites=1}{\Sites}\rbr{1+\conjg{\chikj{\picount}{\sites}}\phikj{\picount}{\sites}}},
\label{eq:insert_integral_bm} \\
c_{\alpha\beta}^{(\picount)}=&\int\frac{\rmd^{2\Sites}\mukj{\picount}{}}{\pi^\Sites}\int\frac{\rmd^{2\Sites}\phikj{\picount}{}}{\pi^\Sites}\exp\rbr{-\abs{\phikv{\picount}}^2-\abs{\mukv{\picount}}^2+\conjg{\phikv{\picount}}\cdot\mukv{\picount}}\conjg{\zetakj{\picount}{\alpha}}\chikj{\picount-1}{\beta}\tbr{\prodl{\sites=0}{\Sites-\max\rbr{\alpha,\beta}-1}\cbr{1+\zetakj{\picount}{\Sites-\sites}\sul{k=1}{\infty}\frac{1}{k!}\rbr{\conjg{\mukj{\picount}{\Sites-\sites}}}^{k}\rbr{\phikj{\picount-1}{\Sites-\sites}}^{k-1}}} \nonumber \\
&\quad\cbr{1+\zetakj{\picount}{\max\rbr{\alpha,\beta}}\sul{k=1}{\infty}c_k^{(2)}\left(\phikj{\picount-1}{\max\rbr{\alpha,\beta}}\right)\left(\conjg{\mukj{\picount}{\max\rbr{\alpha,\beta}}}\right)^k}\tbr{\prodl{\sites=\Sites-\max\rbr{\alpha,\beta}+1}{\Sites-\min\rbr{\alpha,\beta}-1}\cbr{1+\zetakj{\picount}{\Sites-\sites}\sul{k=1}{\infty}c_k^{(3)}\rbr{\phikj{\picount-1}{\Sites-\sites}}\rbr{\conjg{\mukj{\picount}{\Sites-\sites}}}^{k}}} \nonumber \\
&\quad\cbr{1+\zetakj{\picount}{\min\rbr{\alpha,\beta}}\sul{k=1}{\infty}c_k^{(2)}\left(\phikj{\picount-1}{\min\rbr{\alpha,\beta}}\right)\left(\conjg{\mukj{\picount}{\min\rbr{\alpha,\beta}}}\right)^k}\tbr{\prodl{\sites=\Sites-\min\rbr{\alpha,\beta}+1}{\Sites-1}\cbr{1+\zetakj{\picount}{\Sites-\sites}\sul{k=1}{\infty}\frac{1}{k!}\rbr{\conjg{\mukj{\picount}{\Sites-\sites}}}^{k}\rbr{\phikj{\picount-1}{\Sites-\sites}}^{k-1}}}\cbr{\prodl{\sites=1}{\Sites}\rbr{1+\conjg{\chikj{\picount}{\sites}}\phikj{\picount}{\sites}}}
\label{eq:insert_integral_cm}
\end{align}
\begin{align}
d_{\alpha\beta}^{(\picount)}=&\int\frac{\rmd^{2\Sites}\mukj{\picount}{}}{\pi^\Sites}\int\frac{\rmd^{2\Sites}\phikj{\picount}{}}{\pi^\Sites}\exp\rbr{-\abs{\phikv{\picount}}^2-\abs{\mukv{\picount}}^2+\conjg{\phikv{\picount}}\cdot\mukv{\picount}}\cbr{\prodl{\sites=1}{\Sites}\rbr{1+\conjg{\chikj{\picount}{\sites}}\phikj{\picount}{\sites}}}\conjg{\zetakj{\picount}{\alpha}}\conjg{\zetakj{m}{\beta}}\chikj{\picount-1}{\beta}\chikj{\picount-1}{\alpha} \nonumber \\
&\quad\tbr{\prodl{\sites=0}{\Sites-\max\rbr{\alpha,\beta}-1}\cbr{1+\zetakj{\picount}{\Sites-\sites}\sul{k=1}{\infty}\frac{1}{k!}\rbr{\conjg{\mukj{\picount}{\Sites-\sites}}}^{k}\rbr{\phikj{\picount-1}{\Sites-\sites}}^{k-1}}}\cbr{1+\zetakj{\picount}{\max\rbr{\alpha,\beta}}\sul{k=1}{\infty}c_k^{(4)}\left(\phikj{\picount-1}{\max\rbr{\alpha,\beta}}\right)\left(\conjg{\mukj{\picount}{\max\rbr{\alpha,\beta}}}\right)^k} \nonumber \\
&\quad\tbr{\prodl{\sites=\Sites-\max\rbr{\alpha,\beta}+1}{\Sites-\min\rbr{\alpha,\beta}-1}\cbr{1+\zetakj{\picount}{\Sites-\sites}\sul{k=1}{\infty}\frac{1}{k!}\rbr{\conjg{\mukj{\picount}{\Sites-\sites}}}^{k}\rbr{\phikj{\picount-1}{\Sites-\sites}}^{k-1}}}\cbr{1+\zetakj{\picount}{\min\rbr{\alpha,\beta}}\sul{k=1}{\infty}c_k^{(4)}\left(\phikj{\picount-1}{\min\rbr{\alpha,\beta}}\right)\left(\conjg{\mukj{\picount}{\min\rbr{\alpha,\beta}}}\right)^k} \nonumber \\
&\quad\tbr{\prodl{\sites=\Sites-\min\rbr{\alpha,\beta}+1}{\Sites-1}\cbr{1+\zetakj{\picount}{\Sites-\sites}\sul{k=1}{\infty}\frac{1}{k!}\rbr{\conjg{\mukj{\picount}{\Sites-\sites}}}^{k}\rbr{\phikj{\picount-1}{\Sites-\sites}}^{k-1}}},
\label{eq:insert_integral_dm}
\end{align}
with $c_1^{(1)}=c_1^{(2)}=c_1^{(3)}=c_1^{(4)}=1$.

It is important to notice, that the integral over $\phikv{\picount}$ and $\mukv{\picount}$ selects only the $k=1$ terms of the occurring sums. Therefore, the terms with $k\geq2$ can be varied, in order to modify the final path integral in the desired way.

Finally, for $\picount=\pisteps$, a similar argument as for $\picount=0$ allows to restrict the integrals over $\phikv{\pisteps}$ again to those $N_f=\sum_{\sites=1}^{\Sites}\nfj{\sites}$ components with $\nfj{\sites}=1$, while setting all the other components of $\phikv{\pisteps}$ to zero.

After this the $\picount$-th factor in the product over the timesteps only depends on $\zetakv{\picount+1}$ and $\chikv{\picount}$, such that one can easily integrate out the intermediate Grassmann variables $\zetakv{1},\ldots,\zetakv{\pisteps}$ and $\chikv{0},\ldots,\chikv{\pisteps-1}$ by using
\begin{align}
\int&\rmd^{2\Sites}\zeta\intg{}{}{^{2\Sites}\chi}\exp\rbr{-\conjg{\gV{\zeta}}\cdot\gV{\zeta}-\conjg{\gV{\chi}}\cdot\gV{\chi}}\cbr{\prodl{\sites=0}{\Sites-1}\rbr{1+\zeta_{\Sites-\sites}f_{\Sites-\sites}^{(\picount)}}}\cbr{\prodl{\sites=1}{\Sites}\rbr{1+\conjg{\zeta_{\sites}}\chi_{\sites}}}\cbr{\prodl{\sites=1}{\Sites}\rbr{1+\conjg{\chi_{\sites}}\phikj{\picount}{\sites}}}=\prodl{\sites=1}{\Sites}\rbr{1+f_\sites^{(\picount)}\phikj{\picount}{\sites}},
\label{eq:grassman-integrated_nohamiltonian} \\
\int&\rmd^{2\Sites}\zeta\intg{}{}{^{2\Sites}\chi}\exp\rbr{-\conjg{\gV{\zeta}}\cdot\gV{\zeta}-\conjg{\gV{\chi}}\cdot\gV{\chi}}\cbr{\prodl{\sites=0}{\Sites-1}\rbr{1+\zeta_{\Sites-\sites}f_{\Sites-\sites}^{(\picount)}}}\cbr{\prodl{\sites=1}{\Sites}\rbr{1+\conjg{\zeta_{\sites}}\chi_{\sites}}}\cbr{\prodl{\sites=1}{\Sites}\rbr{1+\conjg{\chi_{\sites}}\phikj{\picount}{\sites}}}\conjg{\zeta_{\alpha}}\chi_{\beta}= \nonumber \\
&\qquad f_\alpha^{(\picount)}\phikj{\picount}{\beta}\cbr{\prodl{\sites=1}{\min\rbr{\alpha,\beta}-1}\rbr{1+f_\sites^{(\picount)}\phikj{\picount}{\sites}}}\cbr{\prodl{\sites=\max\rbr{\alpha,\beta}+1}{\Sites}\rbr{1+f_\sites^{(\picount)}\phikj{\picount}{\sites}}}\prodl{\sites=\min\rbr{\alpha,\beta}+1}{\max\rbr{\alpha,\beta}-1}\rbr{1-f_\sites^{(\picount)}\phikj{\picount}{\sites}},
\label{eq:grassman-integrated_sp-part} \\
\int&\rmd^{2\Sites}\zeta\intg{}{}{^{2\Sites}\chi}\exp\rbr{-\conjg{\gV{\zeta}}\cdot\gV{\zeta}-\conjg{\gV{\chi}}\cdot\gV{\chi}}\cbr{\prodl{\sites=0}{\Sites-1}\rbr{1+\zeta_{\Sites-\sites}f_{\Sites-\sites}^{(\picount)}}}\cbr{\prodl{\sites=1}{\Sites}\rbr{1+\conjg{\zeta_{\sites}}\chi_{\sites}}}\cbr{\prodl{\sites=1}{\Sites}\rbr{1+\conjg{\chi_{\sites}}\phikj{\picount}{\sites}}}\conjg{\zeta_{\alpha}}\conjg{\zeta_{\beta}}\chi_{\beta}\chi_{\alpha}= \nonumber \\
&\qquad f_\alpha^{(\picount)}f_\beta^{(\picount)}\phikj{\picount}{\beta}\phikj{\picount}{\alpha}\prodl{\substack{\sites=1 \\ \sites\neq\alpha,\beta}}{\Sites}\rbr{1+f_\sites^{(\picount)}\phikj{\picount}{\sites}}.
\label{eq:grassman-integrated_interaction}
\end{align}
\end{widetext}
Moreover, the integrals over $\zetakv{0}$ and $\chikv{\pisteps}$ yield
\begin{align}
\int\rmd^{2\Sites}\zetakj{0}{}&\exp\rbr{-\conjg{\zetakv{0}}\cdot\zetakv{0}}\cbr{\prodl{\sites=0}{\Sites-1}\rbr{1+\zetakj{0}{\Sites-\sites}\conjg{\mukj{0}{\Sites-\sites}}}} \nonumber \\
&\qquad\prodl{\sites=1}{\Sites}\rbr{\conjg{\zetakj{0}{\sites}}}^{\nij{\sites}}=\prodl{\sites:\nij{\sites}=1}{}\conjg{\mukj{0}{\sites}}
\label{eq:grasmman-integrated_initial} \\
\int\rmd^{2\Sites}\chikj{\pisteps}{}&\exp\rbr{-\conjg{\chikv{\pisteps}}\cdot\chikv{\pisteps}}\cbr{\prodl{\sites=0}{\Sites-1}\rbr{\chikj{\pisteps}{\Sites-\sites}}^{\nfj{\Sites-\sites}}} \nonumber \\
&\qquad\prodl{\sites=1}{\Sites}\rbr{1+\chikj{\pisteps}{\sites}\conjg{\phikj{\pisteps}{\sites}}}=\prodl{\sites:\nfj{\sites}=1}{}\phikj{\pisteps}{\sites}
\label{eq:grasmman-integrated_final}
\end{align}
After performing these integrals, one notices, that the inserted integrals have been chosen such, that the resulting sums can be performed and yield exponentials, such that the propagator is, after integrating out $\mukv{1},\ldots,\mukv{\pisteps}$ as well as $\phikv{0}$ and undo the expansion in $\timestep$, given by the path integral eq.~(\ref{eq:propagator_in_complex_path-integral}), where the classical Hamiltonian is given by

\begin{align}
H&_{cl}\left(\conjg{\gV{\mu}},\gV{\phi}\right)= \nonumber \\
&\sul{\alpha}{}h_{\alpha\alpha}\conjg{\mu_\alpha}\phi_\alpha f_1\left(\conjg{\mu_\alpha},\phi_\alpha\right) \nonumber \\
&+\sul{\substack{\alpha,\beta \\ \alpha\neq\beta}}{}h_{\alpha\beta}\conjg{\mu_\alpha}\phi_\beta f_2\left(\conjg{\mu_\alpha},\phi_\alpha\right)\exp\left(-\conjg{\mu_\beta}\phi_\beta\right){\prod\limits_{l}}^{\alpha,\beta}g\left(\conjg{\mu_l},\phi_l\right) \nonumber \\
&+\sul{\substack{\alpha,\beta \\ \alpha\neq\beta}}{}U_{\alpha\beta}\conjg{\mu_\alpha}\conjg{\mu_\beta}\phi_\alpha\phi_\beta f_3\left(\conjg{\mu_\alpha},\phi_\alpha\right)f_3\left(\conjg{\mu_\beta},\phi_\beta\right),
\label{eq:classical_hamiltonian_mu-phi}
\end{align}
where $f_1$, $f_2$, $f_3$ and $g$ are arbitrary analytic functions satisfying the following conditions:
\begin{align}
f_1\left(0,\phi\right)&=f_2\left(0,\phi\right)=f_3\left(0,\phi\right)=1
\label{eq:constraint_f} \\
g\left(0,\phi\right)&=1
\label{eq:constraint_g_1} \\
\left.\frac{\partial}{\partial\mu^\ast}g\left(\mu^\ast,\phi\right)\right|_{\mu^\ast=0}&=-2\phi.
\label{eq:constraint_g_2}
\end{align}
Moreover, as in section \ref{sec:path_integral}, the product in the third line runs only over those values of $j$, which are lying between $\alpha$ and $\beta$, excluding $\alpha$ and $\beta$ themselves,
\begin{equation}
{\prod\limits_{j}}^{\alpha,\beta}\ldots=\prod\limits_{j=\min\left(\alpha,\beta\right)+1}^{\max\left(\alpha,\beta\right)-1}\ldots
\end{equation}
\section{The semiclassical amplitude}
\label{app:semiclassical_amplitude}
The semiclassical amplitude is given by the integral over the exponential of the second variation of the path integral around the classical path wich can be written as,

\begin{widetext}
\begin{align}
\mathcal{A}_\gamma=\pilimit\frac{1}{\rbr{2\pi}^{2N-1+\rbr{\pisteps-1}\Sites}}&\intg{}{}{^{N-1}\delta\thetakj{0}{}}\intg{}{}{^N\delta\Jkj{\pisteps}{}}\intg{}{}{^N\delta\thetakj{\pisteps}{}}\intg{}{}{^\Sites\delta\Jkj{1}{}}\intg{}{}{^\Sites\delta\thetakj{1}{}}\cdots\intg{}{}{^\Sites\delta\Jkj{\pisteps-1}{}}\intg{}{}{^\Sites\delta\thetakj{\pisteps-1}{}} \nonumber \\
\exp\Bigg\{&-\frac{1}{2}\delta\thetakv{0}\mathbf{P}_i^\prime\pdiff{\phikv{0}}{\thetakv{i}}\cbr{-\exp\cbr{-2\rmi{\rm diag}\rbr{\thetakv{i}}}+\frac{\rmi\timestep}{\hbar}\pdiff{^2{H^{(cl)}}^{(0)}}{{\phikv{0}}^2}}\pdiff{\phikv{0}}{\thetakv{i}}{\mathbf{P}_i^\prime}^{\rm T}\delta\thetakv{0} \nonumber \\
&-\frac{1}{2}\rbr{\begin{array}{c} \delta\thetakv{\pisteps}\mathbf{P}_f \\ \delta\Jkv{\pisteps}\mathbf{P}_f \end{array}}{\mathbf{O}^{(\pisteps)}}^{\rm T}\rbr{\begin{array}{cc}
\exp\cbr{-2\rmi{\rm diag}\rbr{\thetakv{\pisteps}}} & \mathbf{I}_{\Sites} \\ \mathbf{I}_{\Sites} & \frac{\rmi\timestep}{\hbar}\pdiff{^2{H^{(cl)}}^{(\pisteps-1)}}{{\conjg{\phikv{\pisteps}}}^2}
\end{array}}\mathbf{O}^{(\pisteps)}\rbr{\begin{array}{c} \mathbf{P}_f^{\rm T}\delta\thetakv{\pisteps} \\ \mathbf{P}_f^{\rm T}\delta\Jkv{\pisteps} \end{array}} \nonumber \\
&-\frac{1}{2}\sul{\picount=1}{\pisteps-1}\rbr{\begin{array}{c} \delta\thetakv{\picount} \\ \delta\Jkv{\picount} \end{array}}{\mathbf{O}^{(\picount)}}^{\rm T}\rbr{\begin{array}{cc}
\frac{\rmi\timestep}{\hbar}\pdiff{^2{H^{(cl)}}^{(\picount)}}{{\phikv{\picount}}^2} & \mathbf{I}_{\Sites} \\ \mathbf{I}_{\Sites} & \frac{\rmi\timestep}{\hbar}\pdiff{^2{H^{(cl)}}^{(\picount-1)}}{{\conjg{\phikv{\picount}}}^2}
\end{array}}\mathbf{O}^{(\picount)}\rbr{\begin{array}{c} \delta\thetakv{\picount} \\ \delta\Jkv{\picount} \end{array}} \nonumber \\
&+\rbr{\begin{array}{c} \delta\thetakv{1} \\ \delta\Jkv{1} \end{array}}{\mathbf{O}^{(1)}}^{\rm T}\rbr{\begin{array}{c} 0 \\ \mathbf{I}_{\Sites}-\frac{\rmi\timestep}{\hbar}\pdiff{^2{H^{(cl)}}^{(0)}}{\conjg{\phikv{1}}\partial\phikv{0}} \end{array}}\pdiff{\phikv{0}}{\thetakv{i}}{\mathbf{P}_i^\prime}^{\rm T}\delta\thetakv{i} \nonumber \\
&+\rbr{\begin{array}{c} \delta\thetakv{\pisteps}\mathbf{P}_f \\ \delta\Jkv{\pisteps}\mathbf{P}_f \end{array}}{\mathbf{O}^{(\pisteps)}}^{\rm T}\rbr{\begin{array}{cc} 0 & 0 \\ \mathbf{I}_{\Sites}-\frac{\rmi\timestep}{\hbar}\pdiff{^2{H^{(cl)}}^{(\pisteps-1)}}{\conjg{\phikv{\pisteps}}\partial\phikv{\pisteps-1}} & 0 \end{array}}\mathbf{O}^{(\pisteps-1)}\rbr{\begin{array}{c} \delta\thetakv{\pisteps-1} \\ \delta\Jkv{\pisteps-1} \end{array}} \nonumber \\
&+\sul{\picount=1}{\pisteps-2}\rbr{\begin{array}{c} \delta\thetakv{\picount} \\ \delta\Jkv{\picount} \end{array}}{\mathbf{O}^{(\picount)}}^{\rm T}\rbr{\begin{array}{cc} 0 & 0 \\ \mathbf{I}_{\Sites}-\frac{\rmi\timestep}{\hbar}\pdiff{^2{H^{(cl)}}^{(\picount)}}{\conjg{\phikv{\picount+1}}\partial\phikv{\picount}} & 0 \end{array}}\mathbf{O}^{(\picount)}\rbr{\begin{array}{c} \delta\thetakv{\picount} \\ \delta\Jkv{\picount} \end{array}}\Bigg\},
\label{eq:reduced_propagator_phi-phistar-basis}
\end{align}
\end{widetext}
with
\begin{equation}
\mathbf{O}^{(\picount)}=\rbr{\begin{array}{cc}
\pdiff{\phikv{\picount}}{\thetakv{\picount}} & \pdiff{\phikv{\picount}}{\Jkv{\picount}} \\
\pdiff{\conjg{\phikv{\picount}}}{\thetakv{\picount}} & \pdiff{\conjg{\phikv{\picount}}}{\Jkv{\picount}}
\end{array}}.
\label{eq:jacobi-matrix}
\end{equation}
Moreover, ${\rm diag}\rbr{\V{v}}$ is the diagonal $d\times d$-matrix for which the $(j,j)$-th entry is equal to $v_j$, where $d$ is the dimensionality of the vector $\V{v}$ and $\mathbf{P}_{i/f}$ and $\mathbf{P}_{i/f}^\prime$ are defined  as the $N\times\Sites$ and $(N-1)\times\Sites$-matrices, respectively, which project onto the subspace of initially and finally occupied single particle states, with the latter excluding the first occupied one,
\begin{align}
\rbr{\mathbf{P}_{i/f}}_{l\sites}=&\delta_{\sites_{l}^{(\prime)},\sites}
\label{eq:projection_matrices} \\
\rbr{\mathbf{P}_{i/f}^\prime}_{l\sites}=&\delta_{\sites_{l+1}^{(\prime)},\sites},
\label{eq:restricted_projection_matrices}
\end{align}
where $\sites_1<\ldots<\sites_{N}\in\left\{\sites\in\{1,\ldots,\Sites\}:\nij{\sites}=1\right\}$ and $\sites_1^\prime<\ldots<\sites_{N}^\prime\in\left\{\sites\in\{1,\ldots,\Sites\}:\nfj{\sites}=1\right\}$ are the initially, respectively finally, occupied single particle states.

For later reference, we also define $\bar{\mathbf{P}}_{i/f}$ as the complement of $\mathbf{P}_{i/f}$ as well as
\begin{equation}
\mathbf{Q}_{i/f}=\rbr{\begin{array}{c} \bar{\mathbf{P}}_{i/f} \\ \mathbf{P}_{i/f} \end{array}},
\label{eq:reordering_matrix}
\end{equation}
which are the (orthogonal) matrices, which put the components corresponding to initially and finally unoccupied single particle states to the first $\Sites-N$ positions, and those correspondig to occupied single particle states to the last $N$ positions, \textit{i.e.}
\begin{equation}
\mathbf{Q}_{i/f}{\bf n}^{(i/f)}=(\underbrace{0,\ldots,0}_{\Sites-N},\underbrace{1,\ldots,1}_{N})^{\rm T}.
\label{eq:reordering_matrix-action_on_nif}
\end{equation}
The integral over $\delta\thetakv{0}$ is given by

\begin{widetext}
\begin{align}
&\frac{1}{\rbr{2\pi}^{N-1}}\intg{}{}{^{N-1}\delta\thetakj{0}{}}\exp\Bigg\{-\frac{1}{2}\delta\thetakv{0}\mathbf{P}_i^\prime\pdiff{\phikv{0}}{\thetakv{i}}\rbr{-\exp\cbr{-2\rmi{\rm diag}\rbr{\thetakv{i}}}+\frac{\rmi\timestep}{\hbar}\pdiff{^2{H^{(cl)}}^{(0)}}{{\phikv{0}}^2}}\pdiff{\phikv{0}}{\thetakv{i}}{\mathbf{P}_i^\prime}^{\rm T}\delta\thetakv{0} \nonumber \\
&\qquad\qquad\qquad+\rbr{\begin{array}{c} \delta\thetakv{1} \\ \delta\Jkv{1} \end{array}}{\mathbf{O}^{(1)}}^{\rm T}\rbr{\begin{array}{c} 0 \\ \mathbf{I}_{\Sites}-\frac{\rmi\timestep}{\hbar}\pdiff{^2{H^{(cl)}}^{(0)}}{\conjg{\phikv{1}}\partial\phikv{0}} \end{array}}\pdiff{\phikv{0}}{\thetakv{i}}{\mathbf{P}_i^\prime}^{\rm T}\delta\thetakv{i}-\frac{1}{2}\rbr{\begin{array}{c} \delta\thetakv{1} \\ \delta\Jkv{1} \end{array}}{\mathbf{O}^{(1)}}^{\rm T}\rbr{\begin{array}{cc}
\frac{\rmi\timestep}{\hbar}\pdiff{^2{H^{(cl)}}^{(1)}}{{\phikv{1}}^2} & \mathbf{I}_{\Sites} \\ \mathbf{I}_{\Sites} & \frac{\rmi\timestep}{\hbar}\pdiff{^2{H^{(cl)}}^{(0)}}{{\conjg{\phikv{1}}}^2}
\end{array}}\mathbf{O}^{(1)}\rbr{\begin{array}{c} \delta\thetakv{1} \\ \delta\Jkv{1} \end{array}}\Bigg\}= \nonumber \\
&\frac{1}{\sqrt{2\pi}^{N-1}}\tbr{\det\cbr{\mathbf{I}_{\Sites}-\pdiff{^2{H^{(cl)}}^{(0)}}{\rbr{\mathbf{P}_i\phikv{0}}^2}\exp\cbr{2\rmi{\rm diag}\rbr{\thetakv{i}}}}}^{-1}\exp\tbr{-\frac{1}{2}\rbr{\begin{array}{c} \delta\thetakv{1} \\ \delta\Jkv{1} \end{array}}{\mathbf{O}^{(1)}}^{\rm T}\rbr{\begin{array}{cc}
\frac{\rmi\timestep}{\hbar}\pdiff{^2{H^{(cl)}}^{(1)}}{{\phikv{1}}^2} & \mathbf{I}_{\Sites} \\ \mathbf{I}_{\Sites} & \mathbf{X}^{(1)}
\end{array}}\mathbf{O}^{(1)}\rbr{\begin{array}{c} \delta\thetakv{1} \\ \delta\Jkv{1} \end{array}}},
\label{eq:theta0-fluctuation-integral}
\end{align}
where $\mathbf{X}^{(1)}$ is defined as
\begin{equation}
\mathbf{X}^{(1)}=\frac{\rmi\timestep}{\hbar}\pdiff{^2{H^{(cl)}}^{(0)}}{{\conjg{\phikv{1}}}^2}+\rbr{\mathbf{I}_{\Sites}-\frac{\rmi\timestep}{\hbar}\pdiff{^2{H^{(cl)}}^{(0)}}{\conjg{\phikv{1}}\partial\phikv{0}}}{\mathbf{P}_i^\prime}^{\rm T}\tbr{\exp\cbr{-2\rmi{\rm diag}\rbr{\mathbf{P}_i^\prime\thetakv{i}}}-\frac{\rmi\timestep}{\hbar}\pdiff{^2{H^{(cl)}}^{(0)}}{\rbr{\mathbf{P}_i^\prime\phikv{0}}^2}}^{-1}\mathbf{P}_i^\prime\rbr{\mathbf{I}_{\Sites}-\frac{\rmi\timestep}{\hbar}\pdiff{^2{H^{(cl)}}^{(0)}}{\phikv{0}\partial\conjg{\phikv{1}}}}
\label{eq:def_X1_fermions}
\end{equation}
It can be shown, that eq.~(\ref{eq:def_X1_fermions}) can also be written as
\begin{equation}
\mathbf{X}^{(1)}=\frac{\rmi\timestep}{\hbar}\pdiff{^2{H^{(cl)}}^{(0)}}{{\conjg{\phikv{1}}}^2}+\rbr{\mathbf{I}_{\Sites}-\frac{\rmi\timestep}{\hbar}\pdiff{^2{H^{(cl)}}^{(0)}}{\conjg{\phikv{1}}\partial\phikv{0}}}\mathbf{X}^{(0)}\rbr{\mathbf{I}_{\Sites}-\frac{\rmi\timestep}{\hbar}\pdiff{^2{H^{(cl)}}^{(0)}}{{\phikv{0}}^2}\mathbf{X}^{(0)}}^{-1}\rbr{\mathbf{I}_{\Sites}-\frac{\rmi\timestep}{\hbar}\pdiff{^2{H^{(cl)}}^{(0)}}{\phikv{0}\partial\conjg{\phikv{1}}}},
\label{eq:X1_revisited}
\end{equation}
with
\begin{equation}
\mathbf{X}^{(0)}={\mathbf{Q}_i}^{\rm T}\rbr{\begin{array}{cc} 0 \\ & \exp\cbr{2\rmi{\rm diag}\rbr{\mathbf{P}_i^\prime\thetakv{i}}} \end{array}}\mathbf{Q}_i.
\label{eq:X0_def}
\end{equation}
Now, consider the integral
\begin{align}
&\frac{1}{\rbr{2\pi}^{\Sites}}\intg{}{}{^\Sites\delta\Jkj{\picount}{}}\intg{}{}{^\Sites\delta\thetakj{\picount}{}}\exp\Bigg\{-\frac{1}{2}\rbr{\begin{array}{c} \delta\thetakv{\picount+1} \\ \delta\Jkv{\picount+1} \end{array}}{\mathbf{O}^{(\picount+1)}}^{\rm T}\rbr{\begin{array}{cc} \frac{\rmi\timestep}{\hbar}\pdiff{^2{H^{(cl)}}^{(\picount+1)}}{{\phikv{\picount+1}}^2} & \mathbf{I}_{\Sites} \\ \mathbf{I}_{\Sites} & \frac{\rmi\timestep}{\hbar}\pdiff{^2{H^{(cl)}}^{(\picount)}}{{\conjg{\phikv{\picount+1}}}^2} \end{array}}{\mathbf{O}^{(\picount+1)}}\rbr{\begin{array}{c} \delta\thetakv{\picount+1} \\ \delta\Jkv{\picount+1} \end{array}} \nonumber \\
&\qquad-\frac{1}{2}\rbr{\begin{array}{c} \delta\thetakv{\picount} \\ \delta\Jkv{\picount} \end{array}}{\mathbf{O}^{(\picount)}}^{\rm T}\rbr{\begin{array}{cc} \frac{\rmi\timestep}{\hbar}\pdiff{^2{H^{(cl)}}^{(\picount)}}{{\phikv{\picount}}^2} & \mathbf{I}_{\Sites} \\ \mathbf{I}_{\Sites} & X^{(\picount)} \end{array}}{\mathbf{O}^{(\picount)}}\rbr{\begin{array}{c} \delta\thetakv{\picount} \\ \delta\Jkv{\picount} \end{array}}+\rbr{\begin{array}{c} \delta\thetakv{\picount+1} \\ \delta\Jkv{\picount+1} \end{array}}{\mathbf{O}^{(\picount+1)}}^{\rm T}\rbr{\begin{array}{cc} 0 & 0 \\ \mathbf{I}_{\Sites}-\frac{\rmi\timestep}{\hbar}\pdiff{^2{H^{(cl)}}^{(\picount)}}{\conjg{\phikv{\picount+1}}\partial\phikv{\picount}} & 0 \end{array}}{\mathbf{O}^{(\picount)}}\rbr{\begin{array}{c} \delta\thetakv{\picount} \\ \delta\Jkv{\picount} \end{array}} \Bigg\}= \nonumber \\
&\tbr{\det\cbr{\mathbf{I}_{\Sites}-\frac{\rmi\timestep}{\hbar}\pdiff{^2{H^{(cl)}}^{(\picount)}}{{\phikv{\picount}}^2}\mathbf{X}^{(\picount)}}}^{-1}\exp\tbr{-\frac{1}{2}\rbr{\begin{array}{c} \delta\thetakv{\picount+1} \\ \delta\Jkv{\picount+1} \end{array}}{\mathbf{O}^{(\picount+1)}}^{\rm T}\rbr{\begin{array}{cc} \frac{\rmi\timestep}{\hbar}\pdiff{^2{H^{(cl)}}^{(\picount+1)}}{{\phikv{\picount+1}}^2} & \mathbf{I}_{\Sites} \\ \mathbf{I}_{\Sites} & \mathbf{X}^{(\picount+1)} \end{array}}{\mathbf{O}^{(\picount+1)}}\rbr{\begin{array}{c} \delta\thetakv{\picount+1} \\ \delta\Jkv{\picount+1} \end{array}}}
\label{eq:k-th_integration_fermionic_reduced_propagator}
\end{align}
with
\begin{equation}
\mathbf{X}^{(\picount+1)}=\frac{\rmi\timestep}{\hbar}\pdiff{^2{H^{(cl)}}^{(\picount)}}{{\conjg{\phikv{\picount+1}}}^2}+\rbr{\mathbf{I}_{\Sites}-\frac{\rmi\timestep}{\hbar}\pdiff{^2{H^{(cl)}}^{(\picount)}}{\conjg{\phikv{\picount+1}}\partial\phikv{\picount}}}\mathbf{X}^{(\picount)}\rbr{\mathbf{I}_{\Sites}-\frac{\rmi\timestep}{\hbar}\pdiff{^2{H^{(cl)}}^{(\picount)}}{{\phikv{\picount}}^2}\mathbf{X}^{(\picount)}}^{-1}\rbr{\mathbf{I}_{\Sites}-\frac{\rmi\timestep}{\hbar}\pdiff{^2{H^{(cl)}}^{(\picount)}}{\phikv{\picount}\partial\conjg{\phikv{\picount+1}}}}.
\label{eq:recursion_X}
\end{equation}
For $\picount=1$ this is exactly the integral in eq.~(\ref{eq:reduced_propagator_phi-phistar-basis}) after integrating out $\delta\thetakv{0}$ and thus defines $\mathbf{X}^{(2)}$. One then recognizes, that after the $\picount$-th integration, the integral is again of the form of eq.~(\ref{eq:k-th_integration_fermionic_reduced_propagator}) up to the $(\pisteps-1)$-th integration. With this observation, the semiclassical amplitude is given by
\begin{align}
\mathcal{A}_\gamma=&\pilimit\frac{1}{\rbr{2\pi}^{\frac{3N-1}{2}}}\intg{}{}{^N\Jkj{\pisteps}{}}\intg{}{}{^N\thetakj{\pisteps}{}}\prodl{\picount=0}{\pisteps-1}\sqrt{\det\rbr{\mathbf{I}_{\Sites}-\frac{\rmi\timestep}{\hbar}\pdiff{^2{H^{(cl)}}^{(\picount)}}{{\phikv{\picount}}^2}\mathbf{X}^{(\picount)}}}^{-1} \nonumber \\
&\qquad\exp\tbr{-\frac{1}{2}\rbr{\begin{array}{c} \delta\thetakv{\pisteps}\mathbf{P}_f \\ \delta\Jkv{\pisteps}\mathbf{P}_f \end{array}}{\mathbf{O}^{(\pisteps)}}^{\rm T}\rbr{\begin{array}{cc} \exp\cbr{-2\rmi{\rm diag}\rbr{\thetakv{\pisteps}}} & \mathbf{I}_{\Sites} \\ \mathbf{I}_{\Sites} & X^{(\pisteps)} \end{array}}\mathbf{O}^{(\pisteps)}\rbr{\begin{array}{c} \mathbf{P}_f^{\rm T}\delta\thetakv{\pisteps} \\ \mathbf{P}_f^{\rm T}\delta\Jkv{\pisteps} \end{array}}} \nonumber \\
=&\pilimit\frac{1}{\sqrt{2\pi}^{N-1}}\cbr{\prodl{\picount=0}{\pisteps-1}\sqrt{\det\rbr{\mathbf{I}_{\Sites}-\frac{\rmi\timestep}{\hbar}\pdiff{^2{H^{(cl)}}^{(\picount)}}{{\phikv{\picount}}^2}\mathbf{X}^{(\picount)}}}^{-1}}\sqrt{\det\rbr{\mathbf{I}_{N}-\exp\cbr{-2\rmi{\rm diag}\rbr{\mathbf{P}_{f}\thetakv{\pisteps}}}\mathbf{P}_f\mathbf{X}^{(\pisteps)}\mathbf{P}_f^{\rm T}}}^{-1}.
\label{eq:semiclassical_amplitude_preliminary}
\end{align}
\end{widetext}
In the continuous limit, the discrete set of $\mathbf{X}^{(\picount)}$ turns into a function of time $\mathbf{X}(t)$, and (by expanding it up to first order in $\timestep$) is given by eq.~(\ref{eq:deq_X}), and the semiclassical amplitude can be written in the form given in eq.~(\ref{eq:semiclassical_prefactor_intermediately}).
\section{Possible Classical Hamiltonians}
\label{app:classical_hamiltonians}
In this part, we state different possibilities for the classical hamiltonian as can be derived out of similar calculations as in appendix \ref{app:pathintegral} without going furhter into detail.
\subsection{Classical Hamiltonians in the particle picture}
\label{subapp:particle-picture}
First, we present two possibilities arising directly from the derivation presented in appendix \ref{app:pathintegral}, but restrict ourselves to those, which contain $\gV{\mu}$ and $\gV{\phi}$ in a symmetric way and omitting the one already stated in sec.~\ref{sec:path_integral}. These examples shall just illustrate, which kinds of classical Hamiltonians are possible:
\begin{align}
H_{cl}^{(1)}&\left(\conjg{\gV{\mu}},\gV{\phi}\right)= \nonumber \\
&\sul{\alpha}{}h_{\alpha\alpha}\conjg{\mu_\alpha}\phi_\alpha\cos\left(\conjg{\mu_\alpha}\phi_\alpha\right)+\sul{\substack{\alpha,\beta \\ \alpha\neq\beta}}{}U_{\alpha\beta}\conjg{\mu_\alpha}\conjg{\mu_\beta}\phi_\alpha\phi_\beta \nonumber \\
&+\sul{\substack{\alpha,\beta \\ \alpha\neq\beta}}{}h_{\alpha\beta}\conjg{\mu_\alpha}\phi_\beta\exp\left(-\sul{l=\min\left(\alpha,\beta\right)}{\max\left(\alpha,\beta\right)}\conjg{\mu_l}\phi_l\right),
\label{eq:classical_hamiltonian_1}
\end{align}
\begin{align}
H_{cl}^{(2)}&\left(\conjg{\gV{\mu}},\gV{\phi}\right)= \nonumber \\
&\sul{\alpha}{}h_{\alpha\alpha}\conjg{\mu_\alpha}\phi_\alpha\exp\left(\conjg{\mu_\alpha}\phi_\alpha\right) \nonumber \\
&+\sul{\substack{\alpha,\beta \\ \alpha\neq\beta}}{}h_{\alpha\beta}\conjg{\mu_\alpha}\phi_\beta\exp\left(-\conjg{\mu_\beta}\phi_\beta-\conjg{\mu_\alpha}\phi_\alpha\right) \nonumber \\
&\qquad\qquad\times\prodl{l=\min\left(\alpha,\beta\right)+1}{\max\left(\alpha,\beta\right)-1}\left[1-\sinh\left(2\conjg{\mu_l}\phi_l\right)\right] \nonumber \\
&+\sul{\substack{\alpha,\beta \\ \alpha\neq\beta}}{}U_{\alpha\beta}\conjg{\mu_\alpha}\conjg{\mu_\beta}\phi_\alpha\phi_\beta\cosh\left(\conjg{\mu_\alpha}\phi_\alpha\right)\cosh\left(\conjg{\mu_\beta}\phi_\beta\right),
\label{eq:classical_hamiltonian_2}
\end{align}
Next, consider the more general case, that the quantum Hamiltonian is written in the form
\begin{equation}
\hat{H}=\sul{\alpha,\beta}{}h_{\alpha\beta}\hat{c}_\alpha^\dagger\hat{c}_\beta^{}+\sul{\substack{\alpha,\beta,\rho,\nu \\ \alpha\neq\beta,\rho\neq\nu}}{}U_{\alpha\beta\rho\nu}\hat{c}_\alpha^\dagger\hat{c}_\beta^\dagger\hat{c}_\rho^{}\hat{c}_\nu.
\end{equation}
By splitting the interaction term also into (pairwise) diagonal and non-diagonal terms, one can in a similar way as in sec.~\ref{app:pathintegral} construct the following classical Hamiltonian

\begin{widetext}
\begin{align}
H_{cl}\left(\conjg{\gV{\mu}},\gV{\phi}\right)=&\sul{\alpha}{}h_{\alpha\alpha}\conjg{\mu_\alpha}\phi_\alpha f_1\left(\conjg{\mu_\alpha},\phi_\alpha\right)+\sul{\substack{\alpha,\beta \\ \alpha\neq\beta}}{}h_{\alpha\beta}\conjg{\mu_\alpha}\phi_\beta f_2\left(\conjg{\mu_\alpha},\phi_\alpha\right)\exp\left(-\conjg{\mu_\beta}\phi_\beta\right)\prodl{l=\min\left(\alpha,\beta\right)+1}{\max\left(\alpha,\beta\right)-1}g\left(\conjg{\mu_l},\phi_l\right) \nonumber \\
&+\sul{\substack{\alpha,\beta \\ \alpha\neq\beta}}{}U_{\alpha\beta\beta\alpha}\conjg{\mu_\alpha}\conjg{\mu_\beta}\phi_\alpha\phi_\beta f_3\left(\conjg{\mu_\alpha},\phi_\alpha\right)f_3\left(\conjg{\mu_\beta},\phi_\beta\right) \nonumber \\
&+\sul{\substack{\alpha,\beta,\rho \\ \alpha\neq\beta,\rho\neq\alpha,\rho\neq\beta}}{}\cbr{\Theta\rbr{\beta-\alpha}\Theta\rbr{\beta-\rho}+\Theta\rbr{\alpha-\beta}\Theta\rbr{\rho-\beta}-\Theta\rbr{\alpha-\beta}\Theta\rbr{\beta-\rho}-\Theta\rbr{\beta-\alpha}\Theta\rbr{\rho-\beta}} \nonumber \\
&\qquad\qquad\rbr{U_{\alpha\beta\beta\rho}-U_{\alpha\beta\rho\beta}}\conjg{\mu_\alpha}\conjg{\mu_\beta}\phi_\beta\phi_\rho f_1\rbr{\conjg{\mu_\alpha},\phi_\alpha}f_2\rbr{\conjg{\mu_\alpha},\phi_\alpha}\exp\rbr{-\conjg{\mu_\rho}\phi_\rho}\prodl{\sites=\min\rbr{\alpha,\rho}+1}{\max\rbr{\alpha,\rho}-1}g\rbr{\conjg{\mu_\sites},\phi_\sites} \nonumber \\
&+\sul{\substack{\alpha,\beta,\rho \\ \alpha\neq\beta,\rho\neq\alpha,\rho\neq\beta}}{}\cbr{\Theta\rbr{\beta-\alpha}\Theta\rbr{\rho-\alpha}+\Theta\rbr{\alpha-\beta}\Theta\rbr{\alpha-\rho}-\Theta\rbr{\alpha-\beta}\Theta\rbr{\rho-\alpha}-\Theta\rbr{\beta-\alpha}\Theta\rbr{\alpha-\rho}} \nonumber \\
&\qquad\qquad\rbr{U_{\alpha\beta\rho\alpha}-U_{\alpha\beta\alpha\rho}}\conjg{\mu_\alpha}\conjg{\mu_\beta}\phi_\alpha\phi_\rho f_1\rbr{\conjg{\mu_\alpha},\phi_\alpha}f_2\rbr{\conjg{\mu_\beta},\phi_\beta}\exp\rbr{-\conjg{\mu_\rho}\phi_\rho}\prodl{\sites=\min\rbr{\beta,\rho}+1}{\max\rbr{\beta,\rho}-1}g\rbr{\conjg{\mu_\sites},\phi_\sites} \nonumber \\
&+\sul{\substack{\alpha,\beta,\rho,\nu \\ \alpha\neq\beta,\alpha\neq\rho,\alpha\neq\nu,\beta\neq\rho,\beta\neq\nu,\rho\neq\nu}}{}\cbr{\Theta\rbr{\beta-\alpha}-\Theta\rbr{\alpha-\beta}}\cbr{\Theta\rbr{\rho-\nu}-\Theta\rbr{\nu-\rho}}U_{\alpha\beta\rho\nu}\conjg{\mu_\alpha}\conjg{\mu_\beta}\phi_\rho\phi_\nu f_2\rbr{\conjg{\mu_\alpha},\phi_\alpha}f_2\rbr{\conjg{\mu_\beta},\phi_\beta} \nonumber \\
&\qquad\qquad\exp\rbr{-\conjg{\mu_\rho}\phi_\rho-\conjg{\mu_\nu}\phi_\nu}\cbr{\prodl{l=\min\rbr{\alpha,\beta,\rho,\nu}+1}{\min\big\{\tbr{\alpha,\beta,\rho,\nu}\setminus\tbr{\min\rbr{\alpha,\beta,\rho,\nu}}\big\}-1}g\rbr{\conjg{\mu_\sites},\phi_\sites}}\cbr{\prodl{l=\max\big\{\tbr{\alpha,\beta,\rho,\nu}\setminus\tbr{\max\rbr{\alpha,\beta,\rho,\nu}}\big\}+1}{\max\rbr{\alpha,\beta,\rho,\nu}-1}g\rbr{\conjg{\mu_\sites},\phi_\sites}},
\end{align}
\end{widetext}
where $f_1$, $f_2$, $f_3$ and $g$ are again arbitrary analytic functions satisfying eqs.~(\ref{eq:constraint_f}-\ref{eq:constraint_g_2}). Thereby, one should notice, that
\[\min\big\{\tbr{\alpha,\beta,\rho,\nu}\setminus\tbr{\min\rbr{\alpha,\beta,\rho,\nu}}\big\}\]
is the second smallest number out of the set $\tbr{\alpha,\beta,\rho,\nu}$ and
\[\max\big\{\tbr{\alpha,\beta,\rho,\nu}\setminus\tbr{\max\rbr{\alpha,\beta,\rho,\nu}}\big\}\]
the second largest number out of the set $\tbr{\alpha,\beta,\rho,\nu}$.
\subsection{Classical Hamiltonians in the hole picture}
\label{subapp:hole-picture}
The cases considered above, we call particle picture, since the boundary conditions are such, that $\abs{\phi_j}^2=1$ corresponds to the $j$-th single particle state beeing occupied, while $\abs{\phi_j}^2=0$ corresponds to the $j$-th single particle state beeing empty. However, the role of occupied and unoccupied states can be reversed, if eqs.~(\ref{eq:insert_integrals_0}) are replaced by

\begin{widetext}
\begin{align}
&\int\rmd^{2\Sites}\zetakj{0}{}\exp\rbr{-\conjg{\zetakv{0}}\cdot\zetakv{0}}\cbr{\prodl{\sites=1}{\Sites}\rbr{1+\conjg{\chikj{0}{\sites}}\zetakj{0}{\sites}}}\prodl{\sites=1}{\Sites}\rbr{\conjg{\zetakj{0}{\sites}}}^{\nij{\sites}}= \nonumber \\
&\int\frac{\rmd^{2\rbr{\Sites-N_i}}\phikj{0}{}}{\pi^{\Sites-N_i}}\int\limits_{0}^{2\pi}\frac{\rmd^{\Sites-N_i}\thetakj{i}{}}{\rbr{2\pi}^{\Sites-N_i}}\int\rmd^{2\Sites}\zetakj{0}{}\exp\rbr{-\conjg{\zetakv{0}}\cdot\zetakv{0}-\abs{\phikv{0}}^2+\conjg{\phikv{0}}\cdot\mukv{0}}\cbr{\prodl{\sites=1}{\Sites}\rbr{\phikj{0}{\sites}+\conjg{\chikj{0}{\sites}}}}\cbr{\prodl{\sites=0}{\Sites-1}\rbr{\conjg{\mukj{0}{\Sites-\sites}}+\zetakj{0}{\Sites-\sites}}}\prodl{\sites=1}{\Sites}\rbr{\conjg{\zetakj{0}{\sites}}}^{\nij{\sites}},
\label{eq:insert_integrals_0_hole}
\end{align}
where the integrations over $\thetakv{i}$ and $\phikv{0}$ run over those components, which are initially empty $\mukj{0}{\sites}=\rbr{1-\nij{\sites}}\exp\rbr{\rmi\thetakj{i}{\sites}}$, as well as
\begin{align}
a^{(\picount)}=&\int\frac{\rmd^{2\Sites}\mukj{\picount}{}}{\pi^\Sites}\int\frac{\rmd^{2\Sites}\phikj{\picount}{}}{\pi^\Sites}\cbr{\prodl{\sites=1}{\Sites}\rbr{\phikj{\picount}{\sites}}+\conjg{\chikj{\picount}{\sites}}}\exp\rbr{-\abs{\phikv{\picount}}^2-\abs{\mukv{\picount}}^2+\conjg{\phikv{\picount}}\cdot\mukv{\picount}}\prodl{\sites=0}{\Sites-1}\cbr{\sul{k=1}{\infty}\frac{1}{k!}\rbr{\conjg{\mukj{\picount}{\Sites-\sites}}}^{k}\rbr{\phikj{\picount-1}{\Sites-\sites}}^{k-1}+\zetakj{\picount}{\Sites-\sites}},
\label{eq:insert_integral_am_hole} \\
b_{\alpha}^{(\picount)}=&\int\frac{\rmd^{2\Sites}\mukj{\picount}{}}{\pi^\Sites}\int\frac{\rmd^{2\Sites}\phikj{\picount}{}}{\pi^\Sites}\exp\rbr{-\abs{\phikv{\picount}}^2-\abs{\mukv{\picount}}^2+\conjg{\phikv{\picount}}\cdot\mukv{\picount}}\conjg{\zetakj{\picount}{\alpha}}\chikj{\picount-1}{\alpha}\left\{\prodl{\sites=0}{L-\alpha-1}\cbr{\sul{k=1}{\infty}\frac{1}{k!}\rbr{\conjg{\mukj{\picount}{\Sites-\sites}}}^{k}\rbr{\phikj{\picount-1}{\Sites-\sites}}^{k-1}+\zetakj{\picount}{\Sites-\sites}}\right\} \nonumber \\
&\quad\cbr{\sul{k=1}{\infty}c_k^{(1)}\rbr{\phikj{\picount-1}{\alpha}}\rbr{\conjg{\mukj{\picount}{\alpha}}}^{k}+\zetakj{\picount}{\alpha}}\left\{\prodl{\sites=\Sites-\alpha+1}{\Sites-1}\cbr{\sul{k=1}{\infty}\frac{1}{k!}\rbr{\conjg{\mukj{\picount}{\Sites-\sites}}}^{k}\rbr{\phikj{\picount-1}{\Sites-\sites}}^{k-1}+\zetakj{\picount}{\Sites-\sites}}\right\}\cbr{\prodl{\sites=1}{\Sites}\rbr{\phikj{\picount}{\sites}+\conjg{\chikj{\picount}{\sites}}}},
\label{eq:insert_integral_bm_hole} \\
c_{\alpha\beta}^{(\picount)}=&\int\frac{\rmd^{2\Sites}\mukj{\picount}{}}{\pi^\Sites}\int\frac{\rmd^{2\Sites}\phikj{\picount}{}}{\pi^\Sites}\exp\rbr{-\abs{\phikv{\picount}}^2-\abs{\mukv{\picount}}^2+\conjg{\phikv{\picount}}\cdot\mukv{\picount}}\conjg{\zetakj{\picount}{\alpha}}\chikj{\picount-1}{\beta}\tbr{\prodl{\sites=0}{\Sites-\max\rbr{\alpha,\beta}-1}\cbr{\sul{k=1}{\infty}\frac{1}{k!}\rbr{\conjg{\mukj{\picount}{\Sites-\sites}}}^{k}\rbr{\phikj{\picount-1}{\Sites-\sites}}^{k-1}+\zetakj{\picount}{\Sites-\sites}}} \nonumber \\
&\quad\cbr{\sul{k=1}{\infty}c_k^{(2)}\left(\phikj{\picount-1}{\max\rbr{\alpha,\beta}}\right)\left(\conjg{\mukj{\picount}{\max\rbr{\alpha,\beta}}}\right)^k+\zetakj{\picount}{\max\rbr{\alpha,\beta}}}\tbr{\prodl{\sites=\Sites-\max\rbr{\alpha,\beta}+1}{\Sites-\min\rbr{\alpha,\beta}-1}\cbr{\sul{k=1}{\infty}c_k^{(3)}\rbr{\phikj{\picount-1}{\Sites-\sites}}\rbr{\conjg{\mukj{\picount}{\Sites-\sites}}}^{k}+\zetakj{\picount}{\Sites-\sites}}} \nonumber \\
&\quad\cbr{\sul{k=1}{\infty}c_k^{(2)}\left(\phikj{\picount-1}{\min\rbr{\alpha,\beta}}\right)\left(\conjg{\mukj{\picount}{\min\rbr{\alpha,\beta}}}\right)^k+\zetakj{\picount}{\min\rbr{\alpha,\beta}}}\tbr{\prodl{\sites=\Sites-\min\rbr{\alpha,\beta}+1}{\Sites-1}\cbr{\sul{k=1}{\infty}\frac{1}{k!}\rbr{\conjg{\mukj{\picount}{\Sites-\sites}}}^{k}\rbr{\phikj{\picount-1}{\Sites-\sites}}^{k-1}+\zetakj{\picount}{\Sites-\sites}}}\cbr{\prodl{\sites=1}{\Sites}\rbr{\phikj{\picount}{\sites}+\conjg{\chikj{\picount}{\sites}}}}
\label{eq:insert_integral_cm_hole} \\
d_{\alpha\beta}^{(\picount)}=&\int\frac{\rmd^{2\Sites}\mukj{\picount}{}}{\pi^\Sites}\int\frac{\rmd^{2\Sites}\phikj{\picount}{}}{\pi^\Sites}\exp\rbr{-\abs{\phikv{\picount}}^2-\abs{\mukv{\picount}}^2+\conjg{\phikv{\picount}}\cdot\mukv{\picount}}\cbr{\prodl{\sites=1}{\Sites}\rbr{\phikj{\picount}{\sites}+\conjg{\chikj{\picount}{\sites}}}}\conjg{\zetakj{\picount}{\alpha}}\conjg{\zetakj{m}{\beta}}\chikj{\picount-1}{\beta}\chikj{\picount-1}{\alpha} \nonumber \\
&\quad\tbr{\prodl{\sites=0}{\Sites-\max\rbr{\alpha,\beta}-1}\cbr{\sul{k=1}{\infty}\frac{1}{k!}\rbr{\conjg{\mukj{\picount}{\Sites-\sites}}}^{k}\rbr{\phikj{\picount-1}{\Sites-\sites}}^{k-1}+\zetakj{\picount}{\Sites-\sites}}}\cbr{\sul{k=1}{\infty}c_k^{(4)}\left(\phikj{\picount-1}{\max\rbr{\alpha,\beta}}\right)\left(\conjg{\mukj{\picount}{\max\rbr{\alpha,\beta}}}\right)^k+\zetakj{\picount}{\max\rbr{\alpha,\beta}}} \nonumber \\
&\quad\tbr{\prodl{\sites=\Sites-\max\rbr{\alpha,\beta}+1}{\Sites-\min\rbr{\alpha,\beta}-1}\cbr{\sul{k=1}{\infty}\frac{1}{k!}\rbr{\conjg{\mukj{\picount}{\Sites-\sites}}}^{k}\rbr{\phikj{\picount-1}{\Sites-\sites}}^{k-1}+\zetakj{\picount}{\Sites-\sites}}}\cbr{\sul{k=1}{\infty}c_k^{(4)}\left(\phikj{\picount-1}{\min\rbr{\alpha,\beta}}\right)\left(\conjg{\mukj{\picount}{\min\rbr{\alpha,\beta}}}\right)^k+\zetakj{\picount}{\min\rbr{\alpha,\beta}}} \nonumber \\
&\quad\tbr{\prodl{\sites=\Sites-\min\rbr{\alpha,\beta}+1}{\Sites-1}\cbr{\sul{k=1}{\infty}\frac{1}{k!}\rbr{\conjg{\mukj{\picount}{\Sites-\sites}}}^{k}\rbr{\phikj{\picount-1}{\Sites-\sites}}^{k-1}+\zetakj{\picount}{\Sites-\sites}}},
\label{eq:insert_integral_dm_hole}
\end{align}
Inserting the integrals like this results in the following path integral:
\begin{align}
K\left(\V{n}^{(f)},\V{n}^{(i)};t_f\right)=&\left[\prod\limits_{j:n_j^{(i)}=0}^{}\int\limits_{0}^{2\pi}\frac{{\rm d}\theta_j^{(0)}}{2\pi}\exp\left(-{\rm i}\theta_j^{(0)}\right)\right]\left[\prod\limits_{m=1}^{M-1}\prod\limits_{j}^{}\int\limits_{\mathbb{C}}^{}\frac{{\rm d}\phi_j^{(m)}}{\pi}\exp\left(-\left|\phi_j^{(m)}\right|^2\right)\right]\left[\prod\limits_{j:n_j^{(f)}=0}^{}\int\limits_{\mathbb{C}}^{}\frac{{\rm d}\phi_j^{(M)}}{\pi}\phi_j^{(M)}\exp\left(-\left|\phi_j^{(M)}\right|^2\right)\right] \nonumber \\
&\qquad\exp\left\{\sum\limits_{m=1}^{M}\left[{\gV{\phi}^{(m)}}^\ast\cdot\gV{\phi}^{(m-1)}-\frac{{\rm i}\tau}{\hbar}H_{cl}\left({\gV{\phi}^{(m)}}^{\ast},\gV{\phi}^{(m-1)}\right)\right]\right\},
\label{eq:propagator_in_complex_path-integral_hole}
\end{align}
with the classical hamiltonian
\begin{align}
&{H^{(cl)}}^{(\picount)}\rbr{\conjg{\gV{\mu}},\gV{\phi}}= \nonumber \\
&\quad\sul{\alpha=1}{\Sites}h_{\alpha\alpha}^{(\picount)}\exp\rbr{-\conjg{\mu_\alpha}\phi_\alpha}+\sul{\substack{\alpha,\beta=1 \\ \alpha\neq\beta}}{\Sites}U_{\alpha\beta}^{(\picount)}\exp\rbr{-\conjg{\mu_\alpha}\phi_\alpha-\conjg{\mu_\beta}\phi_\beta}+\sul{\substack{\alpha,\beta=1 \\ \alpha\neq\beta}}{\Sites}h_{\alpha\beta}^{(\picount)}\conjg{\mu_\beta}\phi_\alpha\exp\rbr{-\conjg{\mu_\alpha}\phi_\alpha}f\rbr{\conjg{\mu_\beta},\phi_\beta}\prodl{\sites=\min\rbr{\alpha,\beta}+1}{\max\rbr{\alpha,\beta}-1}g\rbr{\conjg{\mu_\sites},\phi_\sites},
\label{eq:classical_hamiltonian_mu-phi_hole}
\end{align}
where $f$ and $g$ are arbitrary analytical functions satisfying
\begin{align}
&f(0,\phi)=1
\label{eq:condition_hole_f} \\
&g(0,\phi)=-1
\label{eq:condition_hole_g_1} \\
&\left.\frac{\partial}{\partial \conjg{\mu}}g\rbr{\conjg{\mu},\phi}\right|_{\conjg{\mu}=0}=2\phi.
\end{align}
\end{widetext}